\documentclass[aps,prd,twocolumn,twoside,superscriptaddress,nofootinbib,10pt]{revtex4-2}
\usepackage{amsmath, amssymb}

\usepackage{silence}
\usepackage{etoolbox}
\usepackage{longtable}
\usepackage{siunitx}
\usepackage{graphicx}
\usepackage{multirow}
\usepackage[dvipsnames]{xcolor}
\usepackage{comment}
\usepackage{orcidlink}
\usepackage{hyperref}
\hypersetup{
  colorlinks,
  citecolor=[HTML]{2d3092},
  linkcolor=[HTML]{2d3092},
  urlcolor=[HTML]{2d3092},
}
\usepackage[capitalise]{cleveref}
\crefname{figure}{Fig.}{Figs.}   \Crefname{figure}{Fig.}{Figs.}
\crefname{equation}{Eq.}{Eqs.}   \Crefname{equation}{Eq.}{Eqs.}
\crefname{table}{Table}{Tables}  \Crefname{table}{Table}{Tables}

\def\dd{\mathrm{d}}
\def\l{\left}
\def\r{\right}
\def\f{\frac}
 
\DeclareSIUnit \parsec {pc}

\newcommand{\new}[1]{\textcolor{black}{{#1}}} 
\newcommand{\hc}[4]{$#1^{+#2}_{-#3}$ (#4\%)}

\makeatletter

\def\@hangfrom@appendix#1#2#3{#1\MakeTextUppercase{\@if@empty{#2}{#3}{#2\@if@empty{#3}{}{:\ #3}}}}

\def\LT@makecaption@rtx#1#2#3{%
  \LT@mcol\LT@cols c{\hbox to\z@{\hss\parbox[t]\LTcapwidth{%
    \sbox\@tempboxa{#1{#2. }#3\unskip\nobreak\vrule\@width\z@\@height\z@\@depth .5\baselineskip}%
    \ifdim\wd\@tempboxa>\hsize #1{#2. }#3\unskip\nobreak\vrule\@width\z@\@height\z@\@depth .5\baselineskip
    \else\hbox to\hsize{\hfil\box\@tempboxa\hfil}\fi\endgraf}\hss}}}

\let\orig@RRAPfmt\frontmatter@RRAP@format
\def\frontmatter@RRAP@format{\vskip 10\p@ \orig@RRAPfmt}   

\def\frontmatter@postabstractspace{30\p@}

\makeatother

\skip\footins=18pt plus 4pt minus 12pt

\BeforeBeginEnvironment{align}{\begingroup\small\thinmuskip=1mu \medmuskip=2mu plus 1mu minus 1mu}
\AfterEndEnvironment{align}{\endgroup}
\BeforeBeginEnvironment{equation}{\begingroup\small}
\AfterEndEnvironment{equation}{\endgroup}

\begin{document}

\title{Enhancing dark siren cosmology via Gaussian process reconstruction \\ of incomplete galaxy catalogs}

\author{Matteo Tagliazucchi\,\orcidlink{0009-0003-8886-3184}}
\email[Contact author: ]{\textcolor{black}{matteo.tagliazucchi2@unibo.it}}
\affiliation{Dipartimento di Fisica e Astronomia ``Augusto Righi''--\href{https://ror.org/01111rn36}{Universit\`{a} di Bologna}, \\
Viale Berti Pichat 6/2, I-40127 Bologna, Italy}
\affiliation{\href{https://ror.org/04j0x0h93}{INFN--Sezione di Bologna}, Viale Berti Pichat 6/2, I-40127 Bologna, Italy}

\author{Jonathan Gair\,\orcidlink{0000-0002-1671-3668}}
\affiliation{\href{https://ror.org/03sry2h30}{Max Planck Institute for Gravitational Physics (Albert Einstein Institute)}, \\
Am M\"{u}hlenberg 1, Potsdam D-14476, Germany}

\author{Riccardo Barbieri\,\orcidlink{0000-0002-4325-054X}}
\affiliation{\href{https://ror.org/03sry2h30}{Max Planck Institute for Gravitational Physics (Albert Einstein Institute)}, \\
Am M\"{u}hlenberg 1, Potsdam D-14476, Germany}

\author{Michele Moresco\,\orcidlink{0000-0002-7616-7136}}
\affiliation{Dipartimento di Fisica e Astronomia ``Augusto Righi''--\href{https://ror.org/01111rn36}{Universit\`{a} di Bologna}, \\
Viale Berti Pichat 6/2, I-40127 Bologna, Italy}
\affiliation{INAF--\href{https://ror.org/00gqsp710}{Osservatorio di Astrofisica e Scienza dello Spazio di Bologna}, \\
via Piero Gobetti 93/3, I-40129 Bologna, Italy}
\affiliation{\href{https://ror.org/04j0x0h93}{INFN--Sezione di Bologna}, Viale Berti Pichat 6/2, I-40127 Bologna, Italy}

\received{20 August 2026}
\accepted[accepted ]{29 September 2026}

\begin{abstract}
We present a novel framework for improving dark siren cosmology by applying a Gaussian process (GP) to the line-of-sight (LOS) reconstruction of incomplete galaxy catalogs.
In the standard galaxy catalog method for inferring the Hubble constant $H_0$ from gravitational-wave (GW) dark sirens, missing galaxies are typically assumed to follow a uniform distribution in comoving volume, an assumption that discards galaxy redshift clustering information crucial for cosmological inference.
We propose instead to model the LOS galaxy redshift distribution as a non-parametric function drawn from a GP realization, which is fitted to the observed incomplete catalog via a hierarchical Bayesian likelihood that explicitly accounts for the survey selection function. 
Applied to mock GW and galaxy catalogs extending up to redshift $z\leq0.4$, our method yields $H_0$ constraints that are on average 23\% more precise than the standard homogeneous completion when using a 24\%-complete galaxy catalog, and 37\% more precise for an 8\%-complete catalog. 
The largest improvement, reaching 66\%, is obtained in configurations where the GP most effectively reconstructs the redshift over- and underdensity features that the homogeneous completion fails to capture. 
\vskip10pt
\noindent DOI: \href{https://doi.org/10.1103/vmzf-1wxq}{10.1103/vmzf-1wxq}
\end{abstract}

\maketitle


\section{Introduction}
The current era of precision cosmology is characterized by a mounting tension in measurements of the Hubble constant $H_0$, which quantifies the present-day expansion rate of the Universe.
On one side, measurements based on the cosmic microwave background under the assumption of the standard $\Lambda$CDM model yield $H_0=67.4\pm0.5\,\si{\kilo\meter\per\second\per\mega\parsec}$ \citep{Planck:2018vyg}, while local distance-ladder measurements using type-Ia supernovae calibrated with Cepheids report $H_0=73.0 \pm 1.0\,\si{\kilo\meter\per\second\per\mega\parsec}$\citep{Riess:2021jrx}.
This discrepancy, now reaching statistical significance exceeding $5\sigma$, is referred to as the Hubble tension and may hint at physics beyond the standard cosmological model \citep{Moresco:2022phi, CosmoVerseNetwork:2025alb}.
To address such tension, novel and independent cosmological probes, which may eventually be combined with each other to reduce systematic effects, are needed.
Gravitational-wave (GW) standard sirens have recently emerged as a new and independent cosmological probe, since their signals encode direct information about the luminosity distance to the source, without requiring any calibration.
By combining these distance measurements with an estimate of the source redshift, the Hubble constant can be inferred using the standard distance-redshift relation.

The redshift information, however, cannot be obtained directly from the GW strain, since the latter is sensitive to the product of the source redshift and chirp mass, making these two perfectly degenerate.
When an unambiguous electromagnetic (EM) counterpart is identified, the host galaxy provides a direct spectroscopic redshift and allows the mass-redshift degeneracy to be broken, as in the bright siren approach \citep{Holz:2005df, Nissanke:2009kt, LIGOScientific:2017vwq, LIGOScientific:2017ync, LIGOScientific:2017adf}.
However, for most detected compact binary coalescences (CBCs)---particularly binary black hole (BBH) mergers---no EM counterpart is expected (and these are therefore called ``dark sirens'').
In the dark siren case, the redshift information may be extracted statistically from the distribution of potential host galaxies within the GW localization volume, as inferred from galaxy catalogs \citep[the ``galaxy catalog method'',][]{Schutz:1986gp, DelPozzo:2011vcw, Chen:2017rfc, LIGOScientific:2018gmd, DES:2020nay, Gair:2022zsa}. 
The galaxy catalog method relies on the assumption that CBCs occur in galaxies, so that the galaxy redshift distribution along the GW line-of-sight (LOS) provides a prior on the source redshift.
In practice, this proceeds as follows: for each GW event, the galaxies within the three-dimensional GW localization volume are identified from a galaxy catalog; their redshift distributions are combined into a redshift prior $p_{\rm cbc}(z)$; and the $H_0$ posterior is obtained by marginalizing the GW likelihood over this prior within a hierarchical Bayesian framework.

To correctly infer the Hubble constant using the galaxy catalog method, it is necessary to properly model the LOS redshift prior, including all possible host galaxies. 
However, all-sky galaxy surveys are affected by selection effects---for example, being magnitude-limited---and therefore cannot detect all galaxies. 
It is thus necessary to correct for galaxy catalog incompleteness to infer $H_0$ without introducing systematic biases \citep{Gray:2019ksv, Perna:2024lod, Hanselman:2024hqy, Borghi:2025pav, VanWyngarden:2025ogy, Alfradique:2025tbj, Cross-Parkin:2025xwf}. 
Indeed, if an incomplete catalog is used directly without accounting for missing galaxies, the cosmological inference effectively downweights redshifts where the catalog is sparse, even if those redshifts are genuinely populated by potential host galaxies. 
The standard approach in current analyses models the contribution from missing galaxies by assuming they are uniformly distributed in comoving volume \citep{LIGOScientific:2018gmd, Finke:2021aom, Gray:2021sew, Gray:2023wgj, Mastrogiovanni:2023emh, Borghi:2023opd, Datrier:2025zdf, LIGOScientific:2026uyd}. 
While this correctly accounts for the overall normalization of the galaxy number density, it implicitly washes out any clustering information present in the data: the over- and underdensities of the galaxy redshift distribution---which are precisely what make the dark siren method informative beyond a prior uniform in comoving volume \citep{Kalomenopoulos:2025qpt}---are diluted in proportion to the catalog incompleteness. 
As a result, at high redshifts, where galaxy catalogs are inevitably highly incomplete, this approach may severely limit the cosmological constraining power of the dark siren method. 
Moreover, in the limit of a large number of GW events, modeling the galaxy catalog incompleteness in this way would inevitably induce a bias, since the assumed LOS redshift prior does not match the true underlying host distribution.

Several alternative approaches have been proposed to improve upon the uniform comoving volume completion. 
For example, \cite{Finke:2021aom} proposes a ``multiplicative completion,'' in which missing galaxies are assumed to trace the distribution of those present in the catalog.
Another strategy incorporates clustering information directly into the incompleteness correction \citep{Dalang:2023ehp, Leyde:2024tov, Dalang:2024gfk, Leyde:2025rzk, Barbieri:2026fdy}. 
In particular, \citet{Leyde:2024tov, Leyde:2025rzk} reconstruct the full galaxy density field---jointly with the galaxy magnitude distribution and detection probability---from a magnitude-limited catalog using Gaussian random fields as nonparametric priors, with a dark matter bias prescription connecting the reconstructed field to the observed galaxies.
\new{An alternative approach proposes to recover missing galaxies using the mean properties of the observed ones to keep clustering information at high redshifts \cite{Zhu:2021aat}.}
Another strategy avoids the incompleteness problem partly by replacing standard galaxy catalogs with more complete alternatives, such as galaxy cluster catalogs or considering only the brightest galaxy subset \citep{Naveed:2025kgk, Beirnaert:2025wcx}. 
Alternatively, GW events can be combined with tracers of the large-scale matter distribution other than galaxies \citep{Scelfo:2021fqe, Dupletsa:2026uqs}. 
A distinct class of methods in which galaxy catalog incompleteness is less critical relies on the angular cross-correlation between GW events and galaxy catalogs \citep{Oguri:2016dgk, Calore:2020bpd, Bera:2020jhx, Mukherjee:2022afz, Ghosh:2023ksl, Ferri:2024amc, Pan:2025iya, Pedrotti:2025tfg, SantiagodeMatos:2025iyj, Cheng:2026atn, Cross-Parkin:2026wyz}.

In this work, we apply Gaussian processes (GPs) to model the LOS redshift prior distribution used in dark siren $H_0$ inference.
Rather than assuming a parametric form for the galaxy distribution or a physically motivated model of galaxy clustering, we represent the unknown host distribution as a nonparametric function, corresponding to the realization of a GP, which specifies a prior distribution over smooth functions.
Given an incomplete galaxy catalog, the GP is fitted to the observed galaxy redshift distribution via a hierarchical Bayesian likelihood that explicitly accounts for the survey selection function, and the resulting posterior over the galaxy distribution is subsequently used as the redshift prior in the $H_0$ inference.
A key advantage of this approach is that the GP framework does not require any modeling of galaxy physics, such as halo occupation distributions, bias models, or the galaxy-matter cross-correlation, all of which carry significant uncertainties at the relevant scales. 
Instead, it learns the correlation structure of the galaxy distribution directly from the data, with the GP kernel parameters marginalizing over the degree of smoothness and the characteristic clustering scale.
This makes the method robust to astrophysical uncertainties while remaining flexible enough to capture the clustering features that are essential for dark siren cosmology.

The paper is organized as follows. 
In \cref{sec:framework} we describe the statistical framework adopted for $H_0$ inference. 
We first introduce the hierarchical Bayesian likelihood for GW events and the simplified GW toy model used throughout the analysis. 
We then describe three approaches to modeling the galaxy redshift distribution considered in this work: the idealized complete catalog case, the standard homogeneous incompleteness correction assuming missing galaxies are uniformly distributed in comoving volume, and the novel GP-based method. 
For the latter, we detail the construction of the galaxy likelihood, the GP prior on the redshift distribution, and the transformation applied to ensure positivity and physical boundedness of the reconstructed distribution. 
In \cref{sec:results} we present the results of the analysis applied to simulated datasets.
We begin by describing the construction of the mock galaxy catalogs and the corresponding mock GW catalogs, generated for a range of detector and survey configurations. 
We then present the GP reconstruction of the galaxy redshift distribution and compare it to the true underlying distribution and to the homogeneous completion, assessing the impact of catalog incompleteness and redshift uncertainties. 
Finally, we apply the reconstructed redshift priors to dark siren cosmology, presenting $H_0$ posterior distributions for all configurations studied.
We conclude in \cref{sec:conclusions} with a summary of our findings and an outlook on future extensions of this work.

\section{Statistical framework}\label{sec:framework}

 In the galaxy catalog method, the likelihood of observing a set of $N_{\rm obs}$ GW events given the Hubble constant $H_0$ can be described by an inhomogeneous Poisson process in the presence of selection effects \cite{Mandel:2018mve, Vitale:2020aaz, Gair:2022zsa}
\begin{equation}\label{eq:hyperlike-scalefree-H0only} 
    \mathcal{L}(\{ \boldsymbol{d}_i \} \!\mid\! H_0) = \prod_{i = 1}^{N_{\rm obs}} \frac{\int \dd z \, \mathcal{L}_{\rm gw}(\boldsymbol{d}_i \!\mid\! d_L(z;H_0)) p_{\rm cbc}(z)}{\int \dd z \, P^{\rm gw}_{\rm det}(z,H_0) \, p_{\rm cbc}(z)}.
\end{equation}
Here, $\{\boldsymbol{d}_i\}$ denotes the data from the observed events. 
The likelihood is evaluated by marginalizing each event's redshift using the CBC redshift prior $p_{\rm cbc}(z)$, while the denominator accounts for GW selection effects. 
The expression in \cref{eq:hyperlike-scalefree-H0only} represents a simplified version of the full hierarchical model \cite{Mandel:2018mve, Vitale:2020aaz, Gair:2022zsa}, since it considers only the luminosity distances inferred from the GW signals, rather than incorporating masses and other waveform parameters, and restricts the inference to $H_0$ alone, treating all remaining population-level hyperparameters as fixed. 
While these simplifications reduce the generality of the model, they allow us to isolate the cosmological information encoded in the distance measurements and the redshift prior, thereby providing a simple and clear framework to introduce the novel methodology presented in the following sections.
The quantity $p_{\rm cbc}(z)$ entering \cref{eq:hyperlike-scalefree-H0only} encodes the redshift distribution of GW sources and may be constructed from galaxy observations, as described in \cref{subsec:gal-modeling}.
We first specify the toy model adopted for the GW likelihood and selection function before turning to the construction of $p_{\rm cbc}(z)$.

\subsection{Gravitational-wave toy model}
 Since we use a simplified hierarchical framework for the inference of $H_0$, we consider a toy model for the single-event GW likelihood that depends only on the observed luminosity distance $d^{\rm obs}_{L,i}$ as
\begin{equation}\label{eq:gw-like-gauss-simplified}
     \mathcal{L}_{\rm gw}(d^{\rm obs}_{L,i} \!\mid\! d_L(z;H_0)) = \f{1}{\sqrt{2\pi} \sigma_{d_L}} e^{- \f{(d^{\rm obs}_{L,i} -d_L(z;H_0))^2}{2\sigma^2_{d_L}}},
\end{equation}
where 
\begin{equation}
    \sigma_{d_L} = A d_L(z;H_0).
\end{equation}
The factor $A$ represents a constant fractional error.

The term $P^{\rm gw}_{\rm det}(z,H_0)$ appearing in \cref{eq:hyperlike-scalefree-H0only} defines the probability of detecting a GW event at redshift $z$ given $H_0$.
It is obtained by integrating the GW likelihood over the space of detectable datasets.
In this simple model, the ``data'' is the observed luminosity distance. We assume that a GW event is detected if this observed luminosity distance falls below a threshold value $d_L^{\rm thr}$---representing the distance horizon of the network of detectors considered:
\begin{align}\label{eq:gw-sel-function-simplified}
    P^{\rm gw}_{\rm det}(z,H_0) & = \int_{-\infty}^{\infty} \dd \hat{d}_L \, \Theta(\hat{d}_L - d_L^{\rm thr})  \mathcal{L}_{\rm gw}(\hat{d}_{L} \!\mid\! d_L(z;H_0)) \nonumber \\
    & = \int_{-\infty}^{d_L^{\rm thr}} \dd \hat{d}_L \, \mathcal{L}_{\rm gw}(\hat{d}_{L} \!\mid\! d_L(z;H_0)) \nonumber \\
    & = \f{1}{2} \l[1 + \mathrm{erf}\l(\f{d_L(z;H_0) - d_L^{\rm thr}}{\sqrt{2}A d_L(z;H_0)}\r)\r].
\end{align}
In the above expression, $\Theta(\cdot)$ is the Heaviside step function and $\mathrm{erf}(\cdot)$ is the error function. 

\begin{figure}
    \centering
    \includegraphics[width=\linewidth]{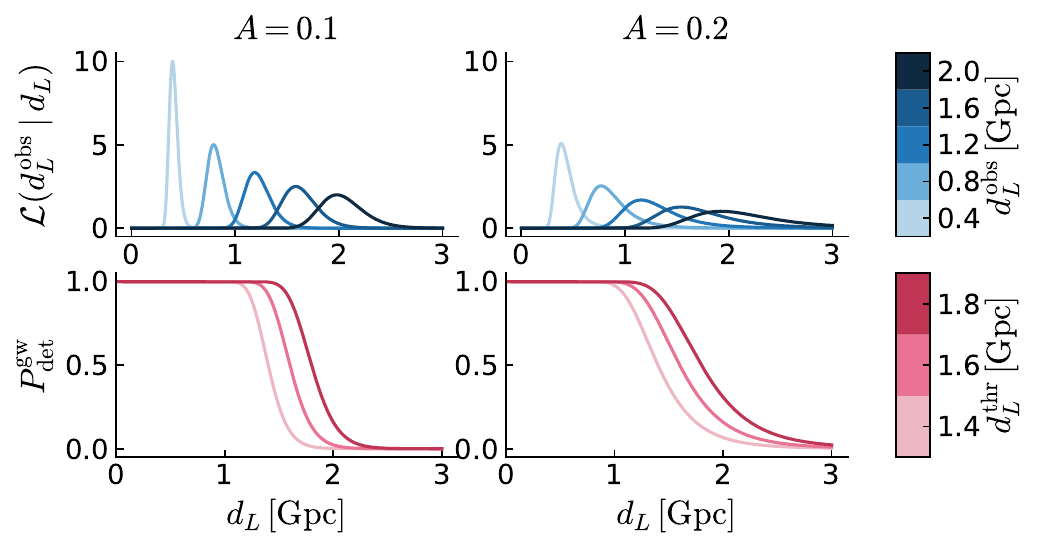}
    \caption{Top: examples of the single-event Gaussian GW likelihood \cref{eq:gw-like-gauss-simplified} as a function of $d_L$ and two choices of the fractional error parameter $A$ (0.1 and 0.2) and for different observed luminosity distances. 
    Bottom: the corresponding detection probability \cref{eq:gw-sel-function-simplified} for the same two values of $A$ and for three different thresholds $d_L^{\rm thr}$.}
    \label{fig:gw-like-and-pdet}
\end{figure}

\Cref{fig:gw-like-and-pdet} shows the GW likelihood and the GW detection probability for different values of $A$ and $d_L^{\rm thr}$.

\subsection{Galaxy catalog modeling and incompleteness correction}\label{subsec:gal-modeling}

The quantity $p_{\rm cbc}(z)$ entering \cref{eq:hyperlike-scalefree-H0only} represents the redshift distribution of the GW sources. 
This distribution is not known \emph{a priori}, but can be constructed from galaxy observations, effectively using the posterior from electromagnetic (EM) observations as a prior for the GW data. 
Throughout the following analyses, we assume that CBC events occur in galaxies, so that $p_{\rm cbc}(z)$ is given by the product of the probability of finding a galaxy at redshift $z$, $p_{\rm gal}(z)$, and the probability that a galaxy at that redshift hosts a GW merger, $p_{\rm rate}(z)$:
\begin{equation}\label{eq:p-cbc}
p_{\rm cbc}(z) = \frac{p_{\rm gal}(z) \, p_{\rm rate}(z)}{\int \dd z \, p_{\rm gal}(z) \, p_{\rm rate}(z)}.
\end{equation}
For simplicity, the merger rate is modeled as a uniform distribution with support in the redshift range $[0,z_{\rm cut}]$, where $z_{\rm cut}$ represents the edge of the mock galaxy catalogs used in the following analyses.
We consider three different approaches to modeling the $p_{\rm gal}$ term.
The first represents an idealized case in which the galaxy catalog is complete.
In the second approach, we account for catalog incompleteness in the way commonly adopted in current analyses \citep{Mastrogiovanni:2023emh, Gray:2023wgj, LIGOScientific:2026uyd}, by assuming that missing galaxies are uniformly distributed in comoving volume.
Both of these approaches are described in this subsection.
The third approach, introduced in \cref{subsec:gp-modeling}, is a novel nonparametric method based on Gaussian processes (GP) that reconstructs the full galaxy distribution from an incomplete catalog while preserving clustering information, which is washed out by the uniform completion but is crucial for dark siren cosmology.

\textbf{\emph{Complete galaxy catalog case.}}
We first consider the case in which the redshift prior is built using a complete galaxy catalog.
In this case, $p_{\rm gal}(z)$ is simply given by the sum of the redshift posterior distributions for all $N_g$ galaxies that fall within the localization volume of the GW event~\footnote{Technically, the sum is over all galaxies in the Universe and the restriction to galaxies in the GW localization volume happens when this is multiplied by the GW likelihood. In practice, it is common to restrict the sum to galaxies that are within the 3D localization volume of the source, for some value of $H_0$ within the prior range, in order to reduce computational costs.}
\begin{equation}\label{eq:pgal-complete}
p_{\rm gal}(z) \equiv p_{\rm cat}(z) = \f{1}{N_g}\sum_{i=1}^{N_g} p(z \!\mid\! \tilde z_i, \sigma_{z,i}),
\end{equation}
where $\tilde z_i$ is the observed galaxy redshift and $\sigma_{z,i}$ is its standard deviation error.
The redshift posterior appearing in the above equation is the product of the galaxy likelihood—assumed to be a Gaussian with mean $\tilde z_i$ and standard deviation $\sigma_{z,i}$—multiplied by a prior distribution and properly normalized.
In the absence of any other galaxy information, the most conservative choice is a prior uniform in comoving volume for the galaxies \citep{Gair:2022zsa}. 
The galaxy redshift posterior is then
\begin{equation}\label{eq:p-z-gal-gaussian}
p(z \!\mid\! \tilde z_i, \sigma_{z,i}) = \f{\mathcal{N}(z; \mu = \tilde z_i, \sigma = \sigma_{z,i}) \, \f{\dd V_c}{\dd z}(z)}{\int \dd z \, \mathcal{N}(z; \mu = \tilde z_i, \sigma = \sigma_{z,i}) \, \f{\dd V_c}{\dd z}(z)}.
\end{equation}
Note that thanks to the proper normalization, this posterior is independent of $H_0$.

\textbf{\emph{Homogeneous incompleteness correction.}}
We next consider the case in which the galaxy survey is affected by selection effects and cannot observe all galaxies.
In this case, the galaxy catalog is incomplete and $p_{\rm gal}(z)$ should be modeled as a sum of a term representing the redshift distribution inferred from the galaxies within the catalog and a term representing the redshift distribution of missing galaxies \citep{Chen:2017rfc, Finke:2021aom}
\begin{equation}
    p_{\rm gal}(z) = f_{\mathcal{R}}  p_{\rm cat}(z) + (1-f_{\mathcal{R}})  p_{\rm miss}(z),
\end{equation}
where $f_{\mathcal R}$ is the average completeness fraction in the localization volume $\mathcal{R}$ of the GW event.
This quantity is computed as the comoving volume-averaged probability that the galaxy survey detects a galaxy at $z$, $P_{\rm det}(z)$:
\begin{equation}
f_{\mathcal{R}} = \f{1}{V_c} \int \dd V_c  P_{\rm det}(z) = \f{1}{V_c} \int \dd z P_{\rm det}(z) \f{\dd V_c}{\dd z}(z).
\end{equation}
  Note that this fraction is $H_0$-independent since both $V_c$ and $\dd V_c / \dd z$ scale as $H_0^{-3}$, but depends on the other cosmological parameters.
The term $p_{\rm cat}$ is the same as in \cref{eq:pgal-complete}, but now the sum extends only over all observed galaxies.
In the homogeneous incompleteness correction, missing galaxies are assumed to be uniformly distributed in comoving volume \citep{Finke:2021aom, Gray:2023wgj, Mastrogiovanni:2023zbw, Borghi:2023opd, Borghi:2025pav}
\begin{equation}
p_{\rm miss}(z) = \f{1-P_{\rm det}(z)}{(1-f_{\mathcal{R}}) V_c}  \f{\dd V_c}{\dd z}(z).
\end{equation}
Therefore, the final expression for the redshift prior in the case of an incomplete galaxy catalog with a ``homogeneous completion'' is
\begin{align}
    p_{\rm gal}(z) = & \l[ \f{1}{V_c} \int \dd z \, P_{\rm det}(z)  \f{\dd V_c}{\dd z}(z) \r] p_{\rm cat}(z) \nonumber \\
    & + \l[1-P_{\rm det}(z)\r] \f{1}{V_c} \f{\dd V_c}{\dd z}(z). \label{eq:pgal-ucv}
\end{align}
While this approach correctly accounts for the overall level of catalog incompleteness, the assumption that unobserved galaxies trace the comoving volume uniformly discards any clustering information present in the data.
As discussed in the following subsection, this limitation motivates the development of a more flexible, nonparametric approach, that will be described in \cref{subsec:gp-modeling}.

\textbf{\emph{Galaxy selection function.}}
The incompleteness correction described previously, as well as the GP-based model that will be detailed in the next section, rely on the galaxy selection function, $P_{\rm det}(z)$.
We model this function in terms of the galaxy absolute magnitudes $M$.
The likelihood of observing a galaxy with magnitude $M$ given parameters $\lambda_\mathcal{G}$ is described by a Schechter function, $\phi(M \!\mid\!\lambda_\mathcal{G})$, normalized over the interval $[M_{\rm min}, M_{\rm max}]$ where it has support \citep{Schechter:1976iz}
\begin{equation}
    \mathcal{L}_{\rm EM}(M\!\mid\!\lambda_{\mathcal{G}}) = \f{\phi(M\!\mid\!\lambda_\mathcal{G})}{\int_{M_{\rm min}}^{M_{\rm max}} \dd M \, \phi(M\!\mid\!\lambda_\mathcal{G})}.
\end{equation}
The Schechter function describes the number density of galaxies in the comoving volume per absolute magnitude.
In the following analyses, we assume a Schechter function, nonevolving in redshift, defined by five parameters $\lambda_{\mathcal{G}}$: a normalization density $\phi_*$, a knee absolute magnitude $M_*$, a faint end $M_{\rm max}$, a bright end $M_{\rm min}$, and a slope parameter $\alpha$. 
Explicitly,
\begin{align}\label{eq:schecther-gp}
    \phi(M\!\mid\!\lambda_\mathcal{G}) & = \l\{\begin{array}{lc}
    0.4 \ln (10) \phi_* x^{\alpha+1} e^{-x}, & M_{\rm min } \leq M \leq M_{\rm max } \\
    0, & \text { otherwise, }
\end{array}\r. \nonumber \\
\text{with}\quad x & = 10^{0.4\l(M_*-M\r)}.
\end{align}
Typically, Schechter parameters are quoted for a Hubble constant of $H_0 = 100\;h\; \si{\kilo\meter\per\second\per\mega\parsec}$.
Rescaling to an arbitrary $h$ follows the relations
\begin{equation}
\begin{aligned}
    M_{\rm min, max, *}(h) & = M_{\rm min, max, *} +  5\log_{10}h \\
    \phi_*(h) & = \phi_* h^3.
\end{aligned}
\end{equation}

The detection probability is obtained by integrating $\mathcal{L}_{\rm EM}(M\!\mid\!\lambda_\mathcal{G})$ over the space of detectable magnitudes. 
A galaxy is considered detectable if its absolute magnitude is brighter than a threshold $M_{\rm thr}(z)$, which follows from the survey’s apparent magnitude limit $m_{\rm thr}$.
This is true for magnitude-limited galaxy surveys, while other surveys may have different and more complex selection functions. 
The conversion between apparent and absolute magnitude thresholds is
\begin{equation}\label{eq:app-mag}
    M_{\rm thr}(z) = m_{\rm thr} - 5 \log_{10} \l( \f{d_L(z\!\mid\!\lambda_c)}{1 \si{\mega\parsec}} \r) - 25,
\end{equation}
so that $M_{\rm thr}(z)$ also scales as $M_{\rm thr}(z) +  5\log_{10}h$.
The EM detection probability is then
\begin{align}\label{eq:pdet-gal-z}
    P_{\rm det}(z) & = \int_{M\in\mathrm{detectable}} \dd M \, \mathcal{L}_{\rm EM}(M\!\mid\!\lambda_{\mathcal{G}})  \nonumber \\ 
    & = \int_{-\infty}^\infty \dd M \, \Theta(M_{\rm thr}(z) - M) \mathcal{L}_{\rm EM}(M\!\mid\!\lambda_{\mathcal{G}}) \nonumber \\ 
    & = \f{\int_{M_{\rm min}}^{M_{\rm thr}(z)} \dd M \phi(M\!\mid\!\lambda_\mathcal{G}) }{\int_{M_{\rm min}}^{M_{\rm max}}\, \dd M \phi(M\!\mid\!\lambda_\mathcal{G})}.
\end{align}
Note that $P_{\rm det}(z)$ reduces to $1$ whenever $M_{\rm thr}(z) < M_{\rm max}$.

\begin{figure}
    \centering
    \includegraphics[width=\linewidth]{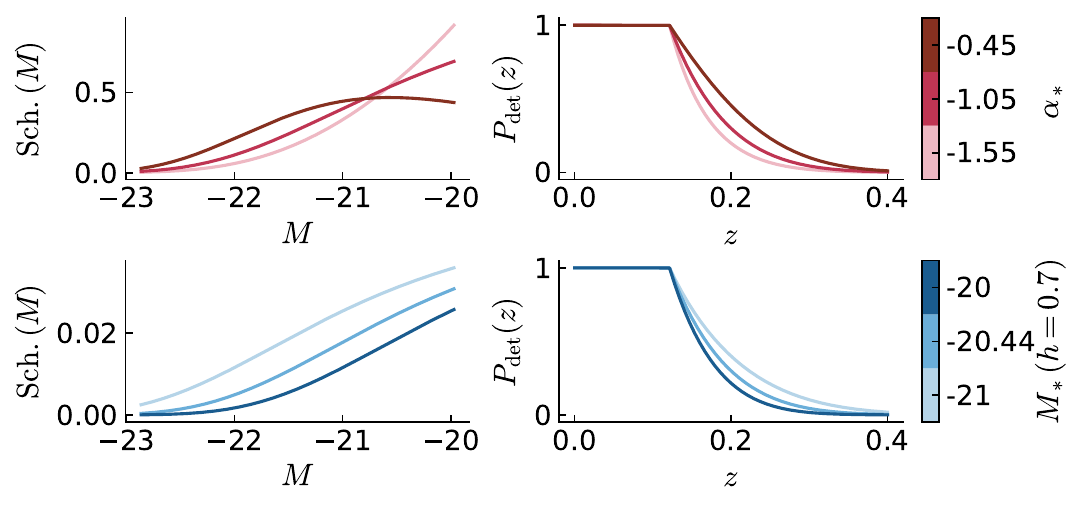}
    \caption{Dependence of the Schechter function and of the corresponding galaxy detection probability on the Schechter parameters $\alpha_*$ (top panels) and $M_*(h=0.7)$ (bottom panels). The other Schechter parameters are fixed to $\phi_*(h=0.7) = 0.0149$, $M_{\rm min}(h=0.7) = -23.29$, and $M_{\rm max}(h=0.7) = -19$.}
    \label{fig:sch-pdet-params}
\end{figure}

Formally, $P_{\rm det}(z)$ depends on the cosmological parameters via the luminosity distance in $M_{\rm thr}(z)$.
In the present analysis, we fix all cosmological parameters except $H_0$.
Under the rescaling of the Schechter parameters, the dependence on $h$ (and thus on $H_0$) cancels.
Indeed, changing the integration variable to $t = M-5\log_{10}h$ yields
\begin{widetext}
\begin{equation}
\resizebox{\textwidth}{!}{$\displaystyle
P_{\rm det}(z) = \f{0.4 \ln (10) \phi_* h^3  \int_{M_{\rm min}+5\log_{10}h}^{M_{\rm thr}(z)+5\log_{10}h} \dd M \, 10^{0.4(\alpha+1)\l[M_*- (M-5\log_{10}h)\r]} e^{-10^{0.4\l[M_*-(M-5\log_{10}h)\r]}}}{0.4 \ln (10) \phi_* h^3  \int_{M_{\rm min}+5\log_{10}h}^{M_{\rm max}+5\log_{10}h} \dd M \, 10^{0.4(\alpha+1)\l[M_*- (M-5\log_{10}h)\r]} e^{-10^{0.4\l[M_*-(M-5\log_{10}h)\r]}}}
= \f{\int_{M_{\rm min}}^{M_{\rm thr}(z)} \dd t \, 10^{0.4(\alpha+1)\l(M_*-t\r)} e^{-10^{0.4\l(M_*-t\r)}}}{\int_{M_{\rm min}}^{M_{\rm max}} \dd t \, 10^{0.4(\alpha+1)\l(M_*-t\r)} e^{-10^{0.4\l(M_*-t\r)}}},$}
\end{equation}
\end{widetext}
which is independent of $H_0$.
Furthermore, we treat the Schechter parameters $\lambda_\mathcal{G}$ as perfectly known in the following, thereby neglecting the dependence of $P_{\rm det}(z)$ on $\lambda_\mathcal{G}$.
In \cref{fig:sch-pdet-params} we show the dependence of the Schechter function \cref{eq:schecther-gp} and of the corresponding galaxy detection probability $P_{\rm det}(z)$ on some of the Schechter parameters.
Note that for more complex or realistic galaxy survey selection functions or more complicated scenarios, the expression for $P_{\rm det}(z)$ must be adjusted accordingly.
For example, for surveys with spatially varying depth or apparent magnitude threshold, $P_{\rm det}$ has to be estimated directly from the data, rather than assumed analytically. 
This could be achieved by fitting a a selection probability to the survey data, or by extending our hierarchical model to include additional selection function parameters.

\subsection{Gaussian process modeling}\label{subsec:gp-modeling}
 The third approach to modeling $p_{\rm gal}(z)$ addresses the main shortcoming of the homogeneous incompleteness correction described above: the loss of galaxy clustering information due to the uniform completion assumption.
Here we present statistical framework aimed at reconstructing the true underlying redshift distribution of galaxies from an incomplete catalog using a nonparametric approach, while retaining the over- and underdensity structure that is crucial for dark siren cosmology.

The likelihood that describes the probability of observing $N^{\rm obs}_{\rm gal}$ galaxies, each with redshift posterior distribution $p(z \!\mid\! \tilde z_i,\sigma_{z,i})$, given the galaxy redshift distribution $p_{\rm gal}$ is (see \cref{app:like-gal-derivation} for a detailed derivation)
\begin{equation}\label{eq:like-gal-gp}
\resizebox{0.85\linewidth}{!}{$\displaystyle
    \mathcal{L}(\{\tilde z_i\}\!\mid\! p_{\rm gal}) = \prod^{N^{\rm obs}_{\rm gal}}_{i=1}  \f{\int \dd z\,  P_{\rm det}(z)\, \mathcal{L}(\tilde z_i \!\mid\! z,\sigma_{z,i})\, p_{\rm gal}(z)}{\int \dd z\, P_{\rm det}(z )\, p_{\rm gal}(z)},$}
\end{equation}
where the single galaxy likelihood $\mathcal{L}(\tilde z_i \!\mid\! z,\sigma_{z,i})$ corresponds to the Gaussian distribution appearing in \cref{eq:p-z-gal-gaussian}.
This likelihood represents an inhomogeneous Poisson process.
Each observed galaxy is treated as a draw from the underlying redshift distribution $p_{\rm gal}(z)$, marginalized over the survey's measurement uncertainty encoded in the single-galaxy likelihood $\mathcal{L}(\tilde z_i \!\mid\! z, \sigma_{z,i})$, modelled as in \cref{eq:p-z-gal-gaussian}.
Each factor in the numerator of the likelihood therefore represents the probability that the $i$th galaxy, whose true redshift is uncertain, is consistent with being drawn from $p_{\rm gal}$.
The term $\alpha(p_{\rm gal}) = \int \dd z P_{\rm det}(z)p_{\rm gal}(z)$ is the fraction of the distribution $p_{\rm gal}$ that lies within the detectable region of the survey, and takes into account that not all galaxies are observable through the detection probability $P_{\rm det}(z)$ defined in \cref{eq:pdet-gal-z}.
This term thus corrects for EM selection effects.
This structure is formally analogous to the likelihood used in GW population inference \cite{Mandel:2018mve, Vitale:2020aaz}, in which one is interested in inferring the parameters $\Lambda$ of a population model $p_{\rm pop}(\theta \!\mid\! \Lambda)$---describing, for instance, the mass and spin distributions of CBCs---given a set of GW detections characterized by source parameters $\theta$.
In this framework, the galaxy redshift distribution $p_{\rm gal}$ plays the role of the population model.

We do not use any particular functional form for $p_{\rm gal}(z)$.
Instead, we adopt a flexible nonparametric model, which allows the shape of the redshift distribution to be determined by the data.
The main motivation is to have a model flexible enough to reconstruct the redshift distribution of the galaxies keeping over- and underdensity information, which are washed out in the "homogeneous completion method" but are crucial for dark siren cosmology, while being simple and not dependent on too many astrophysical assumptions and uncertainties. 
The framework we propose is based on GPs.
A GP is a stochastic process, characterized by a mean function $m(x)$ and a covariance function (or kernel) $K(x,x^\prime)$, that defines a probability distribution over functions
\begin{equation}
    g(x) \sim \mathcal{GP}(m(x), K(x,x^\prime)).
\end{equation}
The defining property of a GP is that for any finite collection of input points $\boldsymbol{x}=\{x_1,\dots,x_n\}$, the marginal distribution of the corresponding function values $g(\boldsymbol{x}) = \{g(x_1),\dots,g(x_n)\}$ is a multivariate Gaussian $
g(\boldsymbol{x}) \sim \mathcal{N}(\boldsymbol{\mu}, \boldsymbol{\Sigma})$, where the mean vector $\boldsymbol{\mu}$ and the covariance matrix $\boldsymbol{\Sigma}$ are defined by the GP mean and covariance functions as $\mu_i = m(x_i)$ and $\Sigma_{ij} = K(x_i,x_j)$, respectively.
Typically, GPs are used in regression problems.
However, here we utilize GP as a prior distribution on the space of functions describing the galaxy distribution $p_{\rm gal}(z)$. 
We choose a mean function and a covariance function, described by some parameters $\lambda_{\rm GP}$, and we build the GP.
We then draw the values of $p_{\rm gal}(z)$ on a defined redshift grid $\boldsymbol{z}=\{z_1,\dots,z_n\}$ from such a GP.
The interpolation of the values $p_{\rm gal}(\boldsymbol{z})$ gives a smooth function modeling the galaxy rate entering in the likelihood \cref{eq:like-gal-gp}.

\begin{figure}[t]
    \centering
    \includegraphics[width=\linewidth]{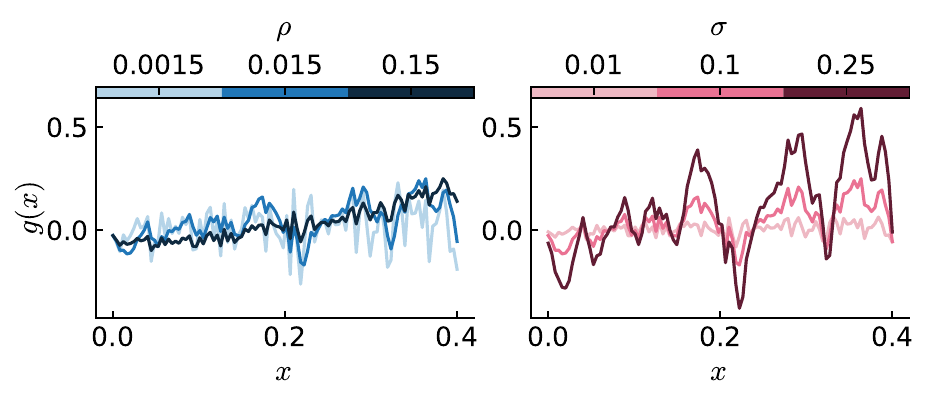}
    \caption{Effects of the parameters $\rho$ and $\sigma$ defining the Matern 5/2 kernel on functions drawn from the corresponding GP. The different curves correspond to different random realizations drawn from the GP prior.}
    \label{fig:gp-kernel-params}
\end{figure}

Rather than fixing $\lambda_{GP}$ to a single value, we effectively marginalize over them. 
To do this, we construct a likelihood for the GP hyperparameters given the observed galaxy catalog and combine it, via Bayes's theorem, with a prior on $\lambda_{\rm GP}$ to obtain a posterior distribution conditioned on the data. 
Marginalizing over this posterior propagates the uncertainty on $\lambda_{\rm GP}$ directly into the distribution of reconstructed GP realizations $p_{\rm gal}(z)$.
This choice encodes very little prior information about the shape of the galaxy redshift distribution, besides enforcing that it must be smooth.

In the following analyses, we consider a zero-mean function. 
The kernel adopted, instead, is the Matern one with $\nu = 5/2$:
\begin{equation}
\begin{aligned}
    K\l(x_i, x_j\r)=\sigma^2\l(1+\f{\sqrt{5} r}{\rho}+\f{5 r^2}{3 \rho^2}\r) \exp \l(-\f{\sqrt{5} r}{\rho}\r),
\end{aligned}
\end{equation}
where $r=\sqrt{\l(x_i-x_j\r)^T\l(x_i-x_j\r)}$.
This kernel was chosen because we checked that it yields results comparable to those obtained with the exponential kernel, which is the standard choice in many problems, while being significantly faster.
This kernel depends on two parameters, $\sigma$ and $\rho$.
The first controls the overall amplitude scale of the covariance, while the latter describes the correlation scale, determining over what input distance the function tends to change significantly.
To see the effects of these kernel parameters and function values drawn from the GP see \cref{fig:gp-kernel-params}.

\begin{figure}[t]
    \centering
    \includegraphics[width=\linewidth]{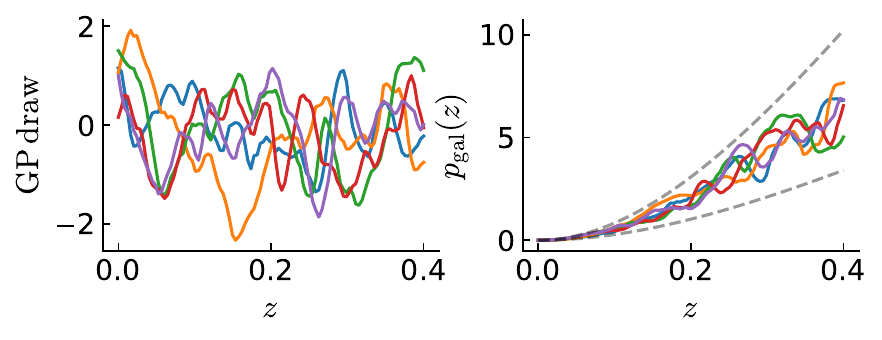}
    \caption{Effect of the transformation \cref{eq:transform-gp} on random GP realizations. 
    Left: random GP realizations.
    Right: random GP realizations transformed using \cref{eq:transform-gp}, compared to the redshift-dependent boundaries \cref{eq:gp-bounds} used in the transformation (dashed lines).}
    \label{fig:gp-transform}
\end{figure}

As shown in \cref{fig:gp-kernel-params}, a function randomly drawn from a GP can take negative values.
Since the galaxy distribution entering in the likelihood \cref{eq:like-gal-gp} must be positive definite, we define $p_{\rm gal}$ as a transformed realization of a GP, using the following transformation:
\begin{equation}\label{eq:transform-gp}
    p_{\rm gal}(z) \propto \mathcal{T}(g(z)) = l(z) + (u(z)-l(z))\cdot\f{1}{1+e^{-g(z)}}.
\end{equation}
The sigmoid function ensures the GP realization is bounded within $(0,1)$.
This bounded output is then scaled to lie between two redshift-dependent boundaries, which scale as the comoving volume element, defined as
\begin{align}
    u(z) & = (1+a) \f{\dd V_c / \dd z}{\int \dd z \, \dd V_c / \dd z} , \nonumber \\
    l(z) & = (1-a) \f{\dd V_c / \dd z}{\int \dd z \, \dd V_c / \dd z} .\label{eq:gp-bounds}
\end{align}
Here, the factor $a$ controls how much the GP realization can spread between the upper and lower boundaries. 
\new{The factor $a$ is therefore crucial to avoid GP overfitting. 
In general, it depends on the number of galaxies and on the redshift errors: we expect fewer fluctuations with many galaxies and large redshift errors, since the sum of many broad galaxy redshift posteriors tends to be smoother. 
Consequently, the value of $a$ in a real survey must be calibrated on simulations, but it is expected to be inversely proportional to the volume of the event considered (a larger volume implies more galaxies, hence a smaller $a$ to avoid overfitting) and to $\sigma_z$ for a similar reason.}
We select boundaries that follow the comoving volume element because the underlying galaxy distribution in the following analyses is expected to be approximately uniform in comoving volume.
This scaling is crucial to speed up the Markov chain Monte Carlo (MCMC) sampling of the likelihood \cref{eq:like-gal-gp}, as described in the following sections.
Note that the comoving volume elements in both boundary functions are normalized, so they no longer depend on $H_0$ and do not artificially introduce information about it.
The effects of such transformations on random GP realizations are shown in \cref{fig:gp-transform}.

\section{Results}\label{sec:results}
 We now apply the Bayesian framework and models described in the previous sections to simulated datasets.
First, we describe the construction of the mock galaxy catalog, as well as the corresponding simulated GW catalogs.
We then present the results of the GP reconstruction of the galaxy redshift distribution. 
Finally, we demonstrate an application to dark siren cosmology.

\subsection{Mock catalogs}

\begin{figure}[b]
    \centering
    \includegraphics[width=\linewidth]{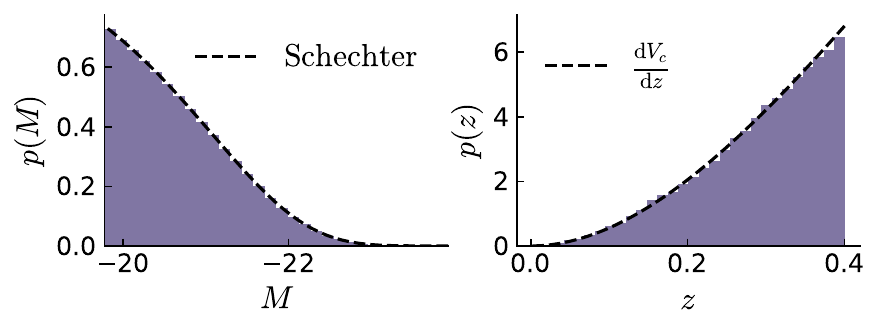}
    \caption{Redshift and absolute magnitude distributions for the MICEv1 mock galaxy catalog subsample (histograms). The black dashed lines show the underlying theoretical redshift distribution (uniform in comoving volume) and the Schechter luminosity function from which the catalog was generated.}
    \label{fig:mice-v1-props}
\end{figure}

\textbf{\emph{Mock galaxy catalogs.}}
The mock galaxy catalogs considered in this work are obtained from a subsample of the MICE grand challenge light-cone simulation (v1) \citep{Carretero:2014ltj, Fosalba:2013wxa, Fosalba:2013mra, Hoffmann:2014ida}.
In fact, since the intrinsic magnitude distribution of the MICE catalog does not match a Schechter function \cref{eq:schecther-gp}, we obtained the subsample via rejection sampling: starting from the full MICE catalog restricted to $z\leq 0.4$, galaxies were accepted or rejected with a probability chosen such that the resulting magnitude distribution follows the Schechter function with fiducial parameters
\begin{align}
    \alpha & = -1.05, \,\, M_{*} = -21.21, \,\, \phi_{*} = 0.0434\,\si{\per\cubic\mega\parsec}, \nonumber \\
    M_{\rm min} & = -24.06, \,\, M_{\rm max} = -19.77.
\end{align}
The resulting subsample contains approximately $3.65 \times 10^6$ galaxies.
\Cref{fig:mice-v1-props} shows the redshift and absolute magnitude distributions of the subsample, comparing them to the theoretical distributions. 
The redshift cut at $z<0.4$ results in a distribution that is uniform in comoving volume for the fiducial MICEv1 cosmology, which assumes a flat $\Lambda$CDM model with $H_0 = 70\,\si{\kilo\meter\per\second\per\mega\parsec}$ and $\Omega_{\rm m,0} = 0.25$.
This redshift cut was chosen precisely to ensure a background galaxy distribution that is approximately uniform in comoving volume, simplifying the modeling of the underlying galaxy population.

\begin{figure}[t]
    \centering
    \includegraphics[width=\linewidth]{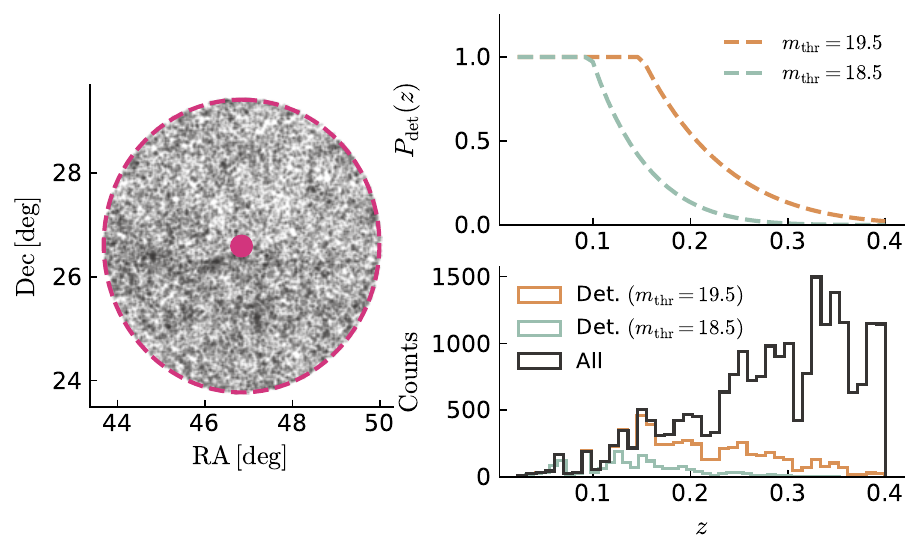}
    \caption{Construction of a single line-of-sight galaxy sample. Left: Schematic of the $25\,\si{\square\deg}$ sky region used for each mock gravitational-wave event. Top right: The galaxy detection probability $P_{\rm det}(z)$ as a function of redshift, resulting from the apparent magnitude cuts $m_{\rm thr}=19.5$ (orange) and $m_{\rm thr}=18.5$ (green). Bottom right: Redshift histogram of the detected galaxies, shown in orange and green for $m_{\rm thr}=19.5$ and $m_{\rm thr}=18.5$, respectively, compared to the true underlying galaxy distribution in the line-of-sight volume (gray).}
    \label{fig:LOS-generation}
\end{figure}

For dark siren cosmology, our goal is to reconstruct the galaxy distribution along a specific LOS corresponding to a GW event's localization area. 
We therefore model each LOS individually.
We consider 1000 different LOS drawn from the original MICE subsample, one for each mock GW event considered in the following analyses.
Specifically, for each mock GW event, the LOS dataset contains all galaxies within a $25\,\si{\square\deg}$ round sky patch centered on the LOS (see left panel of \cref{fig:LOS-generation}). 
This area is comparable to the 90\% localization region of the best-localized BBH event from the O4a observing run \citep{LIGOScientific:2025slb, LIGOScientific:2025jau} and represents the maximum area of the $4\%$ best-localized events in O4b \citep{LIGOScientific:2026uyd, LIGOScientific:2026wfs}. 
It is also representative of the pixel area used when pixelating large localization areas---$500$--$1000\,\si{\square\deg}$---that is the area of roughly $20\%$ of events in the GWTC-5 cosmology analysis \citep{LIGOScientific:2026uyd}.

To simulate realistic survey selection effects, we impose a detection threshold on the apparent magnitude \cref{eq:app-mag}, removing galaxies with $m>m_{\rm thr}$.
We study two cases, $m_{\rm thr} = 19.5,\,18.5$, corresponding to mean completeness fractions (averaged along all LOS) of $24\%$ and $8\%$:
\begin{align}
   m_{\rm thr} = 19.5 & \quad \to \quad \langle P_{\rm compl} \rangle = 24\% ,\nonumber\\
   m_{\rm thr} = 18.5 & \quad \to \quad \langle P_{\rm compl} \rangle = 8\% \label{eq:mag-cut}.
\end{align}
The corresponding detection probabilities $P_{\rm det}(z)$, defined in \cref{eq:pdet-gal-z}, are shown in the top right panel of \cref{fig:LOS-generation}. 
The bottom right panel compares the redshift distribution of the detected galaxies to the true underlying distribution in the considered LOS, illustrating the impact of the magnitude cuts.

For the galaxy redshift likelihood in \cref{eq:p-z-gal-gaussian}, we investigate five scenarios for the redshift uncertainty $\sigma_{z,i}$:
\begin{enumerate}
    \renewcommand{\labelenumi}{(\roman{enumi})}
    \item \texttt{spec-z} (spectroscopic errors): $\f{\sigma_{z,i}}{1+z_i} = 1\times10^{-3}$.
    \item Intermediate case 1: $\f{\sigma_{z,i}}{1+z_i} = 4\times10^{-3}$.
    \item Intermediate case 2: $\f{\sigma_{z,i}}{1+z_i} = 7\times10^{-3}$.
    \item Intermediate case 3: $\f{\sigma_{z,i}}{1+z_i} = 1\times10^{-2}$.
    \item \texttt{DES} (DES survey photometric errors): $\sigma_{z,i} = \min\l[0.013\cdot (1+z_i)^3, \, 0.015\r]$,
\end{enumerate}
where the last case represents photometric redshift uncertainties expected from the Dark Energy Survey \citep[DES][]{DES:2005dhi}, as done in \cite{DES:2019ccw}.

Note that the observed redshift $\tilde z_i$ [the mean $\mu$ in \cref{eq:p-z-gal-gaussian}] is not the true MICEv1 redshift $z_i$, but is instead drawn from a Gaussian distribution centered at the redshift and with standard deviation $\sigma_{z,i}$, effectively perturbing the true value within its measurement uncertainty.
Consequently, larger redshift errors produce a smoother observed distribution. 

\textbf{\emph{Mock GW catalogs.}}
To generate GW catalogs, we assign a binary merger event to each galaxy located in the innermost region of each of the 1000 LOS considered.
We associate events with all galaxies, not only those that are detected, and assign to each GW event the true redshift of its host galaxy.
We then compute the true luminosity distance of each event assuming the fiducial cosmological model used to build the MICEv1 simulation.
To obtain the observed GW luminosity distance, $d_L^{\rm obs}$, we perturb the true luminosity distance by drawing a random value from the Gaussian distribution in \cref{eq:gw-like-gauss-simplified}, centered on the true distance, for a given choice of $A$.
\begin{figure}[t]
    \centering
    \includegraphics[width=\linewidth]{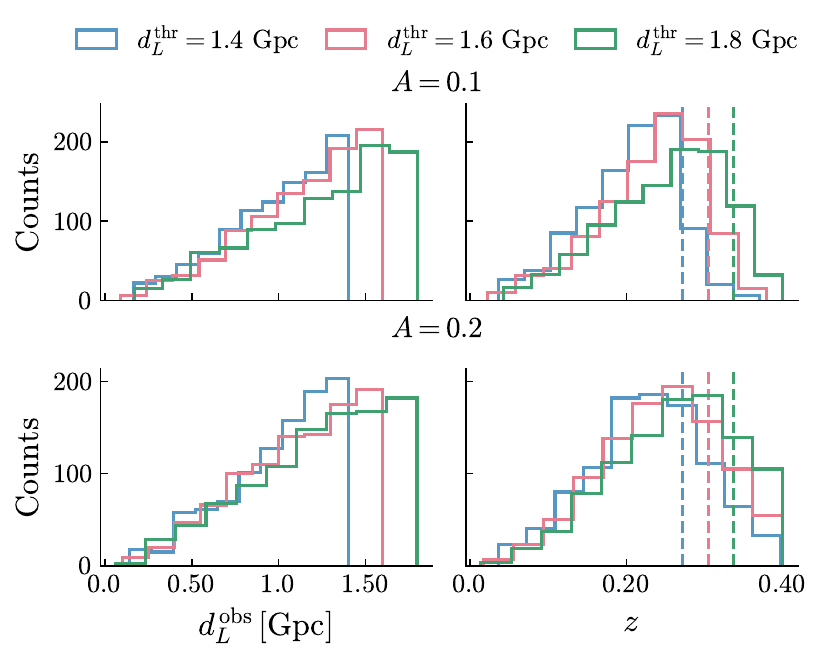}
    \caption{Observed luminosity distance and redshift distributions of the 6 mock GW catalogs generated from 1000 LOS drawn from the MICEv1 subsample.}
    \label{fig:gwmock-cat-gp}
\end{figure}
We consider two configurations defined by $A=0.1$ and $A=0.2$, corresponding to 10\% and 20\% errors on the observed luminosity distance, respectively.
We determine which GW events are detected by applying the criterion $d_L^{\rm obs} < d_{L}^{\rm thr}$, where $d_{L}^{\rm thr}$ is the luminosity distance threshold entering the GW selection function in \cref{eq:gw-sel-function-simplified}.
We consider three luminosity distance thresholds: $d_L^{\rm thr} = 1.4,\,1.6$, and $1.8\,\si{\giga\parsec}$.
For each LOS, among the events satisfying this criterion, we then randomly select one to represent the single GW detection associated with that LOS
This procedure yields six mock GW catalogs of 1000 events each, consistent with the order of magnitude expected from the LIGO-Virgo-KAGRA O5 observing run \citep{KAGRA:2013rdx}.
In \cref{fig:gwmock-cat-gp}, we show the properties of the six catalogs, including their observed luminosity distance distributions and redshift distributions.

\subsection{Galaxy redshift distribution reconstruction}
 We implement the likelihood \cref{eq:like-gal-gp} in \textsc{jax} \citep{jax2018github}, and the construction of the GP is done using the \textsc{tinygp} package \citep{tinygp_software}.
This allows us to sample the likelihood \cref{eq:like-gal-gp} on GPUs using the No-U-Turn \citep[NUTS,][]{hoffman2011nouturnsampleradaptivelysetting} Hamiltonian Monte Carlo sampler implemented in the \textsc{numpyro} framework \citep{phan2019composableeffectsflexibleaccelerated}.
Thanks to GPU acceleration, the fit for each LOS and redshift error assumption takes $\mathcal{O}(10)$ minutes on a single A100 GPU.

\begin{figure*}
    \centering
    \includegraphics[width=\textwidth]{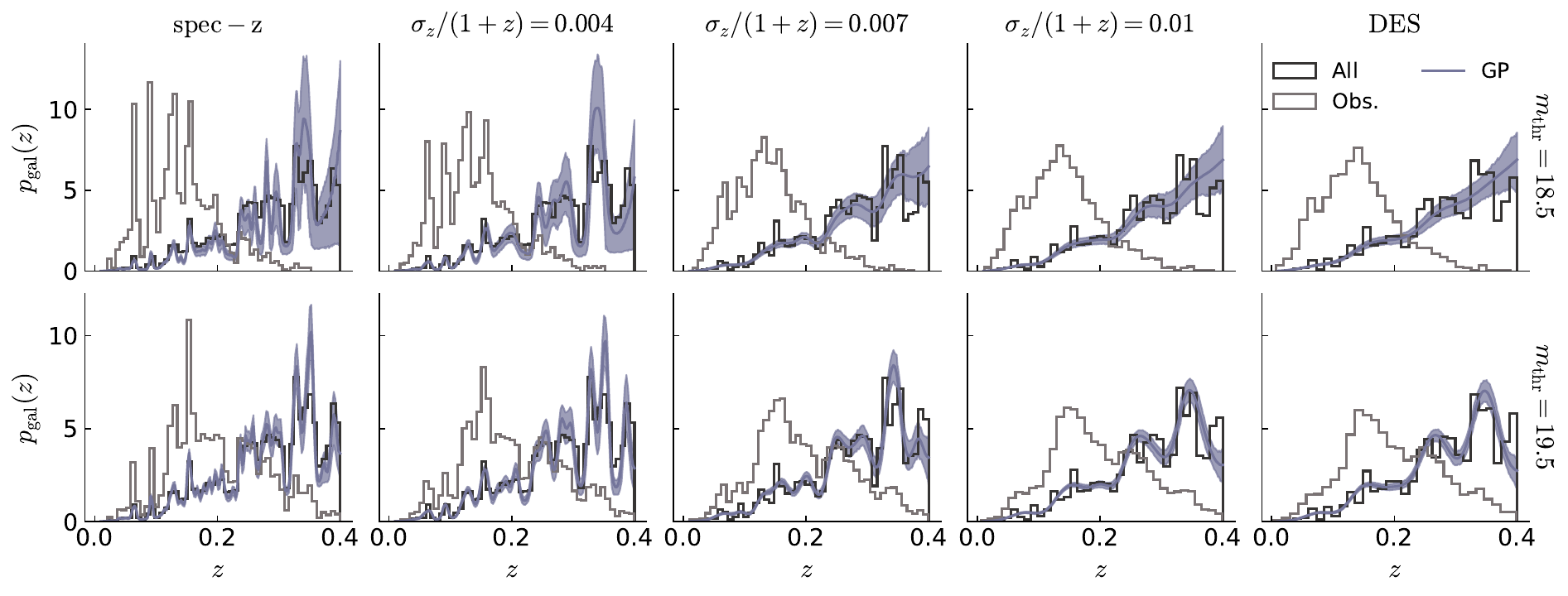}
    \caption{Posterior distribution for $p_{\rm gal}(z)$ obtained with GP model (purple curve) and, for comparison, the true galaxy distribution (black normalized histogram) and observed galaxy distribution (gray normalized histograms). The panels in the first row show results obtained with the catalog cut at $m>m_{\rm thr} = 18.5$, while the second row shows results for $m_{\rm thr} = 19.5$. The results are those relative to the single LOS plotted in \cref{fig:LOS-generation}.}
    \label{fig:pgal-gp-rec}
\end{figure*}
\begin{figure*}
    \centering
    \includegraphics[width=\textwidth]{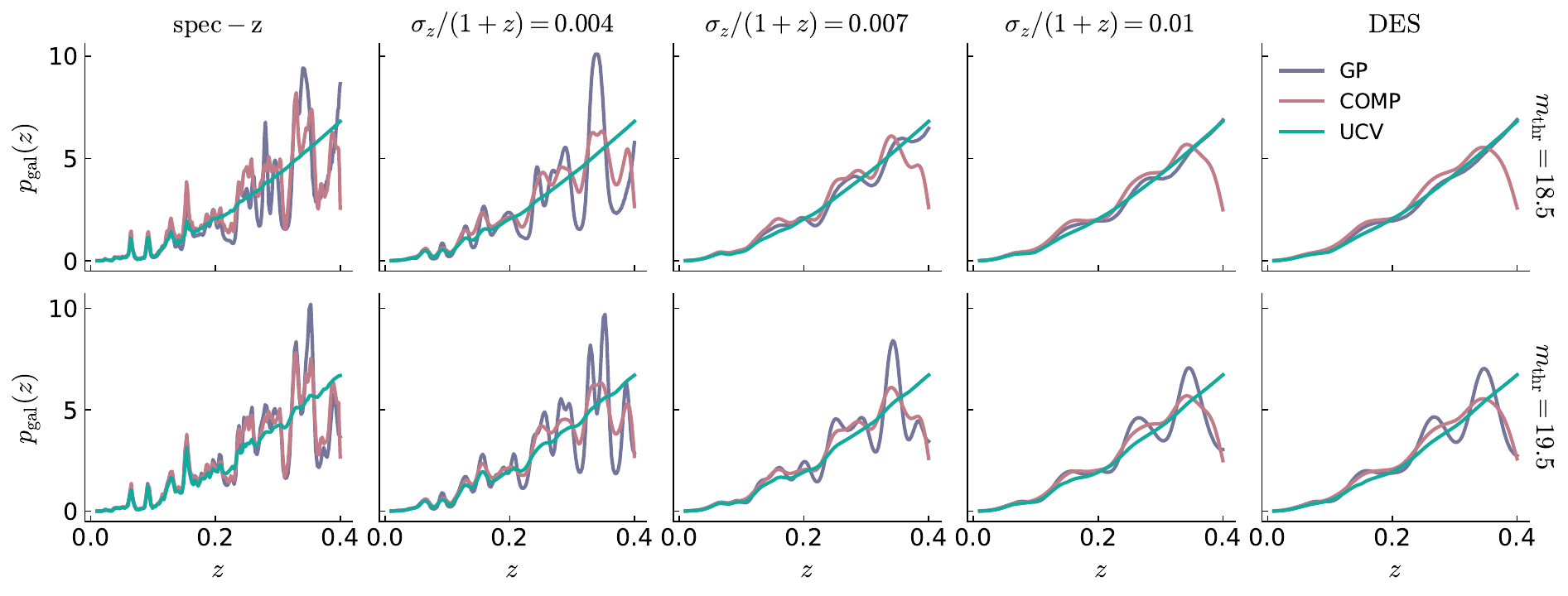}
    \caption{Redshift posterior distributions along the LOS plotted in \cref{fig:LOS-generation} using the catalog with $m_{\rm thr} = 18.5$ (first row) and the one with $m_{\rm thr} = 19.5$ (second row). 
    In each plot, the purple curves correspond to results obtained using the GP-completed galaxy catalog (in particular we show only the median of the GP-reconstructed $p_{\rm gal}(z)$, the green curves to results obtained using the standard homogeneous completion (UCV), and the pink curves to results obtained considering the complete galaxy catalog (COMP.).}
    \label{fig:pgal-coomparison}
\end{figure*}

In \cref{{app:kernel-param-posterior}}, we present and discuss the posterior distribution for the GP kernel parameters obtained with the LOS shown in \cref{fig:LOS-generation}.
In \cref{fig:pgal-gp-rec}, instead, we show the posterior distribution for $p_{\rm gal}(z)$ obtained with the GP method.
In particular, we plot the median and the 68\% credible interval (C.I.).
Top panels show results for the $m_{\rm thr} = 19.5$ catalog, while lower ones for the $m_{\rm thr} = 18.5$ catalog.
The reconstructed $p_{\rm gal}(z)$ follows quite precisely the true underlying galaxy distribution, especially in the $m_{\rm thr} = 19.5$ case.
In the other case, the GP reconstructs fewer features and with larger uncertainty at high redshift, where only a few galaxies are observed.
Nevertheless, some high-redshift features are still reconstructed.
Moreover, the larger the redshift error, the fewer features in the reconstructed redshift distribution are recovered.
The way in which the GP model adds missing galaxies is by adding them where overdensities in the observed redshift distribution are present, effectively adding more galaxies where redshift clusters are present, rather then adding them uniformly in comoving volume as done in the homogeneous completion method.
This behavior is directly tied to the assumed galaxy redshift errors: larger error smooths out the observed overdensities, making them harder for the GP to distinguish from noise, and thus limiting its ability to reconstruct accurately the underlying clustering structure. 
In fact, as it was noted above, larger redshift errors result in fewer recovered features in the reconstructed distribution.

In \cref{fig:pgal-coomparison}, we compare the redshift posterior distributions obtained along the same LOS using three different methods: the GP-completed catalog (purple), the standard homogeneous completion in uniform comoving volume (UCV, green), and the complete galaxy catalog (COMP., pink).
The first row shows results for the $m_{\rm thr} = 18.5$ catalog, while the second row shows results for $m_{\rm thr} = 19.5$.
Overall, the GP-based posteriors are in better agreement with those obtained using the complete catalog than the UCV ones, as the GP method captures the clustering structure of the observed galaxy distribution rather than distributing the missing galaxies uniformly in comoving volume.
In particular, the GP posteriors tend to peak at redshifts where overdensities are present in the observed catalog, closely following the features recovered by the complete catalog.
The UCV method, by contrast, produces smoother and more featureless posteriors, washing out the redshift clustering information.
The improvement brought by the GP completion is more pronounced for the $m_{\rm thr} = 19.5$ catalog, where the denser observed sample allows the GP to reconstruct the underlying galaxy distribution more accurately.
For the shallower $m_{\rm thr} = 18.5$ catalog, the GP posteriors still outperform the UCV ones, although the agreement with the COMP. results is somewhat reduced at high redshift, consistently with the larger reconstruction uncertainty seen in \cref{fig:pgal-gp-rec}.
\new{We note that the median of the reconstructed $p(z)$ can exhibit larger fluctuations than the COMP case, but from \cref{fig:pgal-gp-rec} we note that the COMP distribution is enclosed within the full reconstructed posterior.}

\subsection{Hubble constant constraints}

 Having generated the mock GW catalogs, we can now use the likelihood \cref{eq:hyperlike-scalefree-H0only} to estimate $H_0$.
We model the merger rate distribution $p_{\rm rate}$ entering in \cref{eq:p-cbc} as a uniform distribution in the redshift range $[0, 0.4]$ where the MICEv1 subsample considered has support.
We study three different scenarios described previously to model the $p_{\rm gal}$ distribution.
In the first one the galaxy catalog is 100\% complete and $p_{\rm gal}$, in each LOS, is given by \cref{eq:pgal-complete}, where the observed galaxy redshifts $\tilde z_i$ are obtained by perturbing all true ones using the Gaussian likelihood for galaxy redshifts \cref{eq:p-z-gal-gaussian}.
In the second case, we consider the incomplete galaxy catalog obtained with the apparent magnitude cut \cref{eq:mag-cut} and $p_{\rm gal}$ is built using the homogeneous completion method \cref{eq:pgal-ucv}.
Finally, in the third case we use as $p_{\rm gal}$ the median of the predictive posterior redshift distribution reconstructed from the incomplete catalog using the Gaussian process method.
This choice represents an approximation. 
In principle, one should marginalize over all possible realizations of $p_{\rm gal}$ drawn from each GP. 
Even more consistently, one should perform a joint inference of the redshift distribution and the $H_0$ posterior using both GW and galaxy data simultaneously. 
However, this would require accounting for all events across the $\sim 1000$ LOS, implying the construction and sampling of $\sim 1000$ distinct GP models, which makes the procedure computationally prohibitive for the present analysis.

We study how the $H_0$ posterior changes as a function of the parameters governing both the galaxy observations and the GW likelihood.
Specifically, we vary the apparent magnitude threshold $m_{\rm thr}$ and the redshift uncertainty $\sigma_{z,i}$ to assess the impact of galaxy catalog quality, and the fractional distance error $A$ and luminosity distance threshold $d_L^{\rm thr}$ to probe the sensitivity to GW measurement precision and detector horizon.  
A complete table reporting the median and 68\% C.I. for each configuration studied, is provided in \cref{app:full_H0_posterior}.

\subsubsection{Impact of galaxy survey assumptions}

\begin{figure}[!htb]
    \centering
    \includegraphics[width=\linewidth]{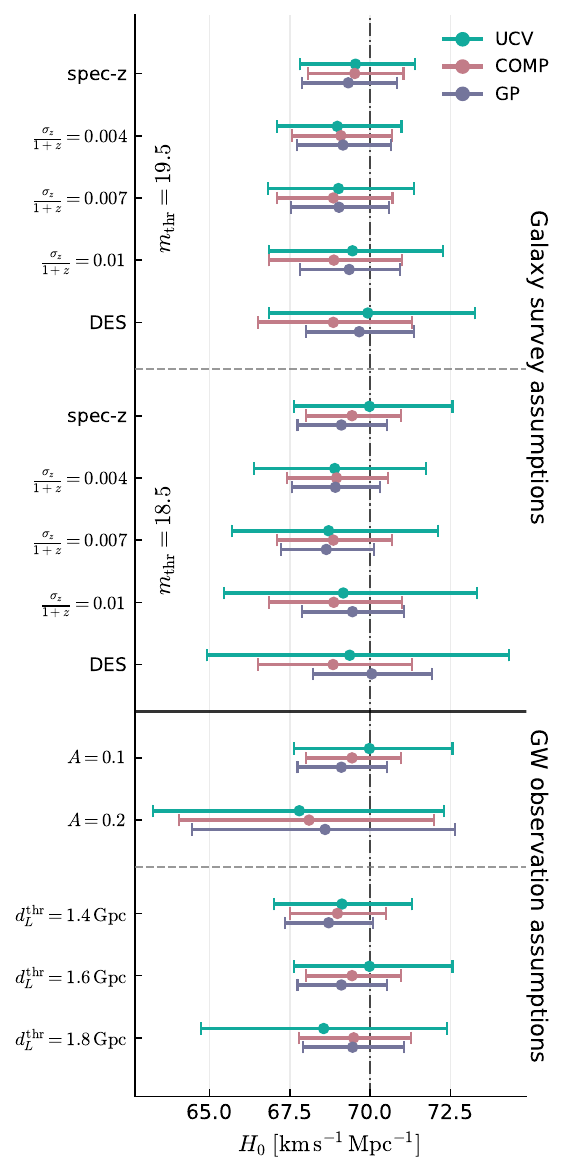}
    \caption{68\% C.I. of the $H_0$ posterior for various configurations, all using the mock GW catalog with $A = 0.1$ and $d_L^{\rm thr} = 1.6\,\si{\giga\parsec}$ unless otherwise stated. The top two panels show results for different redshift errors and magnitude thresholds ($m_{\rm thr} = 19.5$ and $18.5$, respectively). The bottom two panels fix galaxy assumptions to $m_{\rm thr} = 18.5$ and \texttt{spec-z} errors, varying $A$ (third panel) and $d_L^{\rm thr}$ (fourth panel).
    The vertical dashed line marks the injected $H_0$ value.}
    \label{fig:H0-whiskers}
\end{figure}

 We first study how the galaxy survey assumptions affect the $H_0$ posterior.
In the top two panels of \cref{fig:H0-whiskers} we show the $H_0$ 68\% C.I. for various redshift errors and for the two tested apparent magnitude thresholds, $m_{\rm thr} = 18.5$ and $19.5$.
These results are obtained using the mock GW catalog simulated assuming $d_L^{\rm thr} = 1.6\,\si{\giga\parsec}$ and $A = 0.1$, containing 1000 GW events each from a different LOS.

We note that when the redshift error is sufficiently small ($\sigma_z / (1+z) \leq 0.007$), the posterior obtained with the GP method closely matches that from the complete galaxy catalog, while homogeneous completion yields a less informative posterior. 
This reflects the fact that the GP faithfully reconstructs the true underlying galaxy distribution and preserves features at high redshift, as in the complete case, whereas these structures are washed out by homogeneous completion.
When the redshift error becomes larger ($\sigma_z / (1+z) > 0.007$), the GP posterior becomes marginally more informative than even that of the complete galaxy catalog.
This modest effect, not representing a substantial improvement, occurs because the GP method is able to recover some of the true redshift density fluctuations even in the presence of large redshift uncertainties (see \cref{fig:pgal-coomparison}), while in the complete case we effectively marginalize over all galaxies along the LOS. 
Since there are many galaxies along the LOS, each carrying a large redshift uncertainty, the resulting redshift distribution is smoother than the GP one.
Physically, the GP-reconstructed distribution resembles the one obtainable by retaining only clusters of galaxies in redshift space (rather than angular clusters).
This may thus introduce a small bias in regimes where the cluster distribution does not trace well the true host distribution, although we do not find evidence of this in the present analysis.

\subsubsection{Impact of GW observation assumptions}

 We now turn to the impact of GW observation assumptions.
From the last two panels of \cref{fig:H0-whiskers} one can read off the dependence on $d_L^{\rm thr}$ and $A$ at fixed galaxy survey configuration ($m_{\rm thr} = 18.5$ and \texttt{spec-z} redshift errors).

For the GP and complete-catalog methods, the constraints remain broadly comparable when increasing $d_L^{\rm thr}$ from $1.4$ to $1.8\,\si{\giga\parsec}$, with only a mild degradation in some configurations. 
By contrast, the homogeneous completion method becomes significantly less informative at large $d_L^{\rm thr}$.
This is because increasing the GW horizon probes higher-redshift regions where the galaxy catalog is much more incomplete and the homogeneous completion increasingly dominates the reconstructed redshift prior, washing out the clustering information relevant for dark siren cosmology. 
The GP method instead remains able to partially reconstruct the underlying overdensity structure even in these sparsely sampled regions, making it substantially more precise despite the large incompleteness at high redshift.
As expected, the $A=0.1$ configurations systematically yield tighter constraints than the $A=0.2$ ones, since smaller fractional luminosity-distance uncertainties translate directly into sharper GW likelihoods and therefore more informative $H_0$ posteriors.

\subsubsection{Summary of results}
 On average across all simulations tested, the GP method provides constraints that are $23\%$ more precise than those obtained with the homogeneous completion method when using the $m_{\rm thr} = 19.5$ galaxy catalog (see \cref{tab:H0-constraints-summary-gp}).
The largest improvement, $52\%$, is obtained in the \texttt{DES} case with $A=0.1$ and $d_L^{\rm thr} = 1.8\,\si{\giga\parsec}$.
When using the $8\%$ complete galaxy catalog ($m_{\rm thr} = 18.5$) instead, the improvements are much larger, averaging $37\%$.
The maximum improvement, $66\%$, is obtained in the $\sigma_z/(1+z)=0.007$ case with $A=0.1$ and $d_L^{\rm thr} = 1.8\,\si{\giga\parsec}$.
This is expected: with a shallower catalog the homogeneous completion reconstructs redshift features much less accurately, especially at high redshift where observed galaxies are sparse, whereas the GP completion is still able to recover some of the high-redshift structure and thus provide tighter constraints on $H_0$.

\begin{figure}
    \centering
    \includegraphics[width=\linewidth]{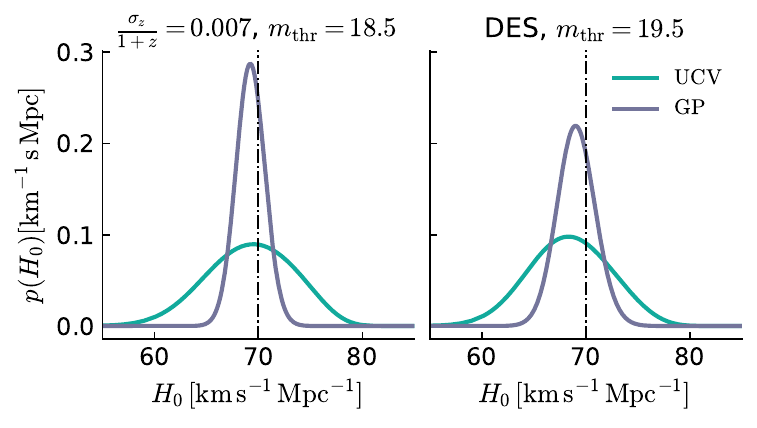}
    \caption{$H_0$ posterior distributions for the two configurations exhibiting the largest GP improvement over homogeneous completion: $\sigma_z/(1+z) = 0.007$ with $m_{\rm thr} = 18.5$ (left) and \texttt{DES}-like redshift errors with$m_{\rm thr} = 19.5$ (right), both at $A = 0.1$ and $d_L^{\rm thr} = 1.8\,\si{\giga\parsec}$.}
    \label{fig:H0-best_gp}
\end{figure}

\Cref{fig:H0-best_gp} illustrates the two configurations that yield the largest GP improvement over UCV: the $\sigma_z/(1+z) = 0.007$, $m_{\rm thr} = 18.5$ case (left panel) and the \texttt{DES}, $m_{\rm thr} = 19.5$ case (right panel), both at $A = 0.1$ and $d_L^{\rm thr} = 1.8\,\si{\giga\parsec}$. 
In both cases, the largest improvement is achieved at the maximum luminosity distance threshold considered, $d_L^{\rm thr} = 1.8\,\si{\giga\parsec}$, where a relevant fraction of the observed GW events lie at redshifts where the galaxy catalog is most incomplete. 
For these events, homogeneous completion assumes a smooth redshift prior, neglecting the clustering features that are instead partially recovered by the GP method, which consequently yields markedly tighter constraints on $H_0$.

\section{Conclusions}\label{sec:conclusions}
 
We have presented a novel method for improving $H_0$ inference from GW dark sirens by applying Gaussian processes directly to the reconstruction of the line-of-sight galaxy redshift distribution from incomplete catalogs. 
Our approach addresses a fundamental limitation of the standard homogeneous incompleteness correction, which assumes that galaxies missing from a catalog are uniformly distributed in comoving volume. 
While this assumption correctly accounts for the overall normalization of the galaxy number density, it washes out the over- and underdensity structure of the galaxy distribution, that is the information that drives the constraining power in the dark siren method. 
In contrast, the GP framework reconstructs the true underlying distribution nonparametrically, preserving redshift clustering features and learning the characteristic correlation scale of the galaxy distribution directly from the data, without requiring any explicit galaxy bias or halo occupation model.

We validated the method on mock datasets derived, considering a range of galaxy survey completeness levels, redshift uncertainty assumptions, and GW detector configurations. 
In all cases, the GP method yields $H_0$ constraints consistent with the true fiducial value and systematically more precise than those from homogeneous completion, with improvements averaging 23\% and 37\% for galaxy catalogs with mean completeness of 24\% and 8\%, respectively. 
The GP reconstruction is particularly valuable at large luminosity distance thresholds, where the catalog is most incomplete and the homogeneous completion is least informative.

Several directions for future work naturally follow from this analysis. 
First, the GP framework could be extended to incorporate galaxy weighting,  assigning each galaxy a weight proportional to its expected merger rate---for example based on luminosity or stellar mass---as done in \cite{Borghi:2025pav}. 
Second, the present analysis approximates the full Bayesian inference by using only the median of the GP-reconstructed redshift distribution when computing $H_0$ posteriors. 
A fully Bayesian treatment, in which the redshift distribution along each LOS and $H_0$ are inferred jointly, marginalizing over all GP realizations simultaneously, remains to be developed.
Since this is prohibitive from the computational point of view, dedicated strategies have to be developed. 
\new{Finally, the method should be applied to real GW and galaxy catalog data, to assess its performance on real observations.
In this context, it will be crucial to include the mixing of spectroscopic and photometric redshifts, so that the GP can exploit the precise low-redshift measurements to better reconstruct the LOS redshift prior at high redshift, where photometric uncertainties dominate, ultimately providing improved constraints on $H_0$.}

\section*{Acknowledgments}
 
M.~T. thanks the Max Planck Institute for Gravitational Physics--Albert Einstein Institute (Potsdam) for the hospitality while this work has been developed.
M.~M. acknowledges the financial contribution from the Grant No. PRIN-MUR 2022 2022NY2ZRS 001 “Optimizing the extraction of cosmological information from Large Scale Structure analysis in view of the next large spectroscopic surveys” and from the ASI Grant No. 2024-10-HH.0 “Attività scientifiche per la missione Euclid–fase E”. 
J.~G. acknowledges support from the Simons Foundation International via Grant No. SFI-MPS- BH-00012593-06. 
We also acknowledge ISCRA for awarding the projects GPGW (No. HP10CF55AE, PI: M.~T.) and MLGW (HP10CM1UJX, PI: M.~T.) access to the LEONARDO supercomputer, owned by the EuroHPC Joint Undertaking, hosted by CINECA (Italy).

\section*{Data availability}
There are no publicly available research data or software supporting this manuscript. Requests for further information or data should be sent to the authors.

\appendix

\section{Galaxy likelihood derivation}\label{app:like-gal-derivation}

In this section, we derive the likelihood \cref{eq:like-gal-gp} used to reconstruct the true underlying redshift distribution of galaxies from an incomplete catalog using a nonparametric approach.
The derivation follows the same hierarchical approach used for gravitational-wave population inference \citep{Loredo:2004nn, Mandel:2018mve, Vitale:2020aaz, Gair:2022zsa}.
We are interested in the probability of observing $N^{\rm obs}_{\rm gal}$ galaxies with redshifts $\{z_i\}$, given some function $\mu(z) \equiv \f{\dd N_{\rm gal}}{\dd z\,}$ describing the true underlying galaxy population rate.
Such a function can be parametrized by parameters $\Lambda$ or approximated using nonparametric approaches.
To be as general as possible, we write a likelihood for $\mu$ rather than for its parameters $\Lambda$.

We start by discretizing the redshift axis into fine bins of width $\delta z_k$, chosen such that each bin contains at most one detected galaxy.
In each bin centered at $z_k$, the number of galaxies is modeled as a Poisson random variable with rate $\mu(z_k) \delta z_k$. 
The probability of observing $\nu$ galaxies in the $k$th bin is
\begin{equation}\label{eq:like-gal-poiss}
    p(\nu \!\mid\! \mu(z_k)) 
    = e^{-\delta z_k \mu(z_k)} \f{\l(\delta z_k \mu(z_k)\r)^\nu}{\nu!}.
\end{equation}
The full likelihood for the observed set $\{z_i\}$ is the product of two terms: 
the probability of detecting exactly one galaxy in each of the $N^{\mathrm{obs}}_{\mathrm{gal}}$ bins containing only one detected galaxy ($p_1$), and the probability of detecting zero galaxies in all other bins ($\bar{p}$):
\begin{equation}\label{eq:like-gal-factorized}
    \mathcal{L}(\{z_i\}\!\mid\! \mu) = \prod_{i=1}^{N^{\rm obs}_{\rm gal}} p_1(\delta z_i) \prod_{k\neq i} \bar{p}(\delta z_k).
\end{equation}

\paragraph{Zero-detection bins.}
Let $P_{\rm det}(z)$ denote the probability of detecting a galaxy at redshift $z$.
For a bin with center $z_k$, the probability of zero detections is obtained by marginalizing over the true (unknown) number of galaxies $\nu$ in that bin---given by \cref{eq:like-gal-poiss}, each weighted by the probability that not one of them is detected
\begin{align}
    \bar{p}(\delta z_k) & = \sum_{\nu = 0}^{\infty} e^{-\delta z_k \mu(z_k)} \f{\l(\delta z_k \mu(z_k)\r)^\nu}{\nu!}\l(1-P_{\rm det}(z_k  )\r)^\nu \nonumber \\
    & = e^{-\delta z_k \mu(z_k)}  \sum_{\nu = 0}^{\infty} \f{1}{\nu!}\l[\delta z_k \mu(z_k)\l(1-P_{\rm det}(z_k  )\r)\r]^\nu \nonumber \\
    & = e^{-\delta z_k \mu(z_k) P_{\rm det}(z_k )}.
\end{align}
The product over all empty bins---which is the second term in \cref{eq:like-gal-factorized}---therefore becomes 
\begin{equation}\label{eq:like-gal-2nd}
    \prod_{k\neq i} \bar{p}(\delta z_k) = e^{-\sum_{k\neq i} \delta z_k \mu(z_k) P_{\rm det}(z_k )}.
\end{equation}

\paragraph{Single-detection bins.}
For a bin centered at $z_i$ containing one detected galaxy, we must marginalize over the true number of galaxies $\nu \geq 1$, with exactly one of them detected and the remaining $\nu-1$ missed.
The number of ways to choose which of the $\nu$ galaxies is the detected one is $\nu$, and the probability of that specific configuration (one detected, $\nu-1$ missed) is $P_{\rm det}(z_i)\l(1-P_{\rm det}(z_i)\r)^{\nu-1}$:
\begin{align}\label{eq:like-gal-p1}
    p_1(\delta z_i) & = \sum_{\nu = 1}^{\infty} \nu\, P_{\rm det}(z_i) \, e^{-\delta z_i \mu(z_i)} \f{\l(\delta z_i \mu(z_i)\r)^\nu}{\nu!}\l(1-P_{\rm det}(z_i  )\r)^{\nu-1} \nonumber \\
    & = P_{\rm det}(z_i)\, e^{-\delta z_i \mu(z_i)} \delta z_i \mu(z_i) \nonumber \\
    & \quad \times \sum_{\nu = 1}^{\infty} \f{\l[\delta z_i \mu(z_i)\l(1-P_{\rm det}(z_i  )\r)\r]^{\nu-1}}{(\nu-1)!} \nonumber \\
    & = \delta z_i \mu(z_i)\, P_{\rm det}(z_i)\, e^{-\delta z_i \mu(z_i) P_{\rm det}(z_i  )},
\end{align}
where the additional factor $\nu$ is due to the fact that there are $\nu$ ways to choose which galaxy is detected, and the factor $P_{\rm det}(z_i)$ is the probability that the chosen galaxy is indeed the one detected.
The product over all detected galaxies---the first term in \cref{eq:like-gal-factorized}---is then 
\begin{align}
    \prod_{i=1}^{N^{\rm obs}_{\rm gal}} p_1(\delta z_i) = & (\delta z_i)^{N^{\rm obs}_{\rm gal}} e^{-\sum^{N^{\rm obs}_{\rm gal}}_{i=1} \delta z_i \mu(z_i) P_{\rm det}(z_i )} \nonumber \\ & \times \prod^{N^{\rm obs}_{\rm gal}}_{i=1} \mu(z_i)\, P_{\rm det}(z_i) \label{eq:like-gal-1st}.
\end{align}

\paragraph{Continuum limit.}
Substituting \cref{eq:like-gal-1st,eq:like-gal-2nd} into \cref{eq:like-gal-factorized} and taking the continuum limit $\delta z \to 0$ (so that sums become integrals) yields
\begin{equation}\label{eq:like-gal-continuum}
    \mathcal{L}(\{z_i\}\!\mid\! \mu) = e^{-\int \dd z\, \mu(z) P_{\rm det}(z )} \prod^{N^{\rm obs}_{\rm gal}}_{i=1} \mu(z_i)\, P_{\rm det}(z_i),
\end{equation}
where we absorbed the constant factor $(\delta z_i)^{N^{\rm obs}_{\rm gal}}$ into an overall normalization. Note that \cref{eq:like-gal-continuum} is the standard result for a thinned Poisson process: if the true process has intensity $\mu(z)$ and each point is independently retained with probability $P_{\rm det}(z)$, the observed process has intensity $\mu(z)P_{\rm det}(z)$, both for the exponential (survival) term and for the density of observed points.

\paragraph{Including galaxy redshift errors.}
If the galaxy catalog provides not exact redshifts but only noisy measurements $\tilde z_i$ for each galaxy (with measurement likelihood $\mathcal{L}(\tilde z_i \!\mid\! z, \sigma_{z,i})$), the likelihood must be marginalized over the true unknown redshift $z$ of each detected galaxy. Since $P_{\rm det}(z)$ is a property of the true redshift $z$ (through the true apparent magnitude of the galaxy) and not of the noisy measurement, it must remain inside this marginalization together with $\mu(z)$:
\begin{align}\label{eq:gal-like-rate-errors}
    \mathcal{L}(\{\tilde z_i\}\!\mid\! \mu) = & e^{-\int \dd z\, \mu(z) P_{\rm det}(z)} \nonumber \\ 
    & \times \prod^{N^{\rm obs}_{\rm gal}}_{i=1} \int \dd z\, \mu(z)\, P_{\rm det}(z)\, \mathcal{L}(\tilde z_i \!\mid\! z, \sigma_{z,i}) ,
\end{align}
where $\mathcal{L}(\tilde z_i \!\mid\! z, \sigma_{z,i})$ is modeled as in \cref{eq:p-z-gal-gaussian}, now interpreted as the measurement likelihood rather than a posterior on $z$. We stress that this assumes detection depends only on the true redshift $z$ and not on the redshift measurement error, i.e. that whether a galaxy is detected is unaffected by the (photometric or spectroscopic) redshift uncertainty $\sigma_{z,i}$; this is a reasonable assumption since detection is set by apparent magnitude, which is determined by the true redshift.

\paragraph{Scale-free likelihood.}
We are interested in inferring the galaxy distribution $p_{\rm gal}$, rather than the full galaxy rate $\mu$.
To write the likelihood for $p_{\rm gal}$ instead of $\mu$, we factorize the latter as
\begin{equation}
    \mu(z) \equiv \f{\dd N_{\rm gal}}{\dd z}= N\, p_{\rm gal}(z).
\end{equation}
Then, we marginalize \cref{eq:gal-like-rate-errors} over $N$ assuming a scale-free prior, that is $\pi(\log(N)) = \mathrm{Unif}$:
\begin{widetext}
\begin{align}\label{eq:gpgal-marg-N}
    \mathcal{L}(\{\tilde z_i\}\!\mid\! p_{\rm gal}) & = \int_0^\infty \f{\dd N}{N} \, \pi(N) e^{- N \int \dd z\, p_{\rm gal}(z) P_{\rm det}(z )} \prod^{N^{\rm obs}_{\rm gal}}_{i=1} \int \dd z\, N p_{\rm gal}(z)\, P_{\rm det}(z)\, \mathcal{L}(\tilde z_i \!\mid\! z,\sigma_{z,i}) \nonumber \\
    & = \l( \prod^{N^{\rm obs}_{\rm gal}}_{i=1} \int \dd z\, p_{\rm gal}(z)\, P_{\rm det}(z)\, \mathcal{L}(\tilde z_i \!\mid\! z,\sigma_{z,i})\r) \int_0^\infty \f{\dd N}{N} \, N^{N^{\rm obs}_{\rm gal}} e^{- N \int \dd z\, p_{\rm gal}(z) P_{\rm det}(z )} \nonumber \\
    & = \l(\f{1}{(\alpha(p_{\rm gal}))^{N^{\rm obs}_{\rm gal}}} \prod^{N^{\rm obs}_{\rm gal}}_{i=1} \int \dd z\, p_{\rm gal}(z)\, P_{\rm det}(z)\, \mathcal{L}(\tilde z_i \!\mid\! z,\sigma_{z,i})\r)  \int_0^\infty \dd t \, t^{N^{\rm obs}_{\rm gal} - 1} e^{-t},
\end{align}
\end{widetext}
where we used the integration variable $t = N \alpha(p_{\rm gal})$ and we defined
\begin{equation}
    \alpha(p_{\rm gal}) = \int \dd z\, p_{\rm gal}(z) P_{\rm det}(z ),
\end{equation}
representing the fraction of galaxies in the population that are expected to be detected.
The last integral in \cref{eq:gpgal-marg-N} is simply equal to a constant factor $(N_{\rm gal}^{\rm obs}-1)!$ and can thus be omitted.
The final expression for the scale-free likelihood is then
\begin{equation}
    \mathcal{L}(\{\tilde z_i\}\!\mid\! p_{\rm gal}) \propto \prod^{N^{\rm obs}_{\rm gal}}_{i=1} \f{\int \dd z\, p_{\rm gal}(z)\, P_{\rm det}(z)\, \mathcal{L}(\tilde z_i \!\mid\! z,\sigma_{z,i})}{\alpha(p_{\rm gal})}.
\end{equation}

\section{Kernel parameter constraints}\label{app:kernel-param-posterior}

\begin{figure*}
    \centering
    \includegraphics[width=\linewidth]{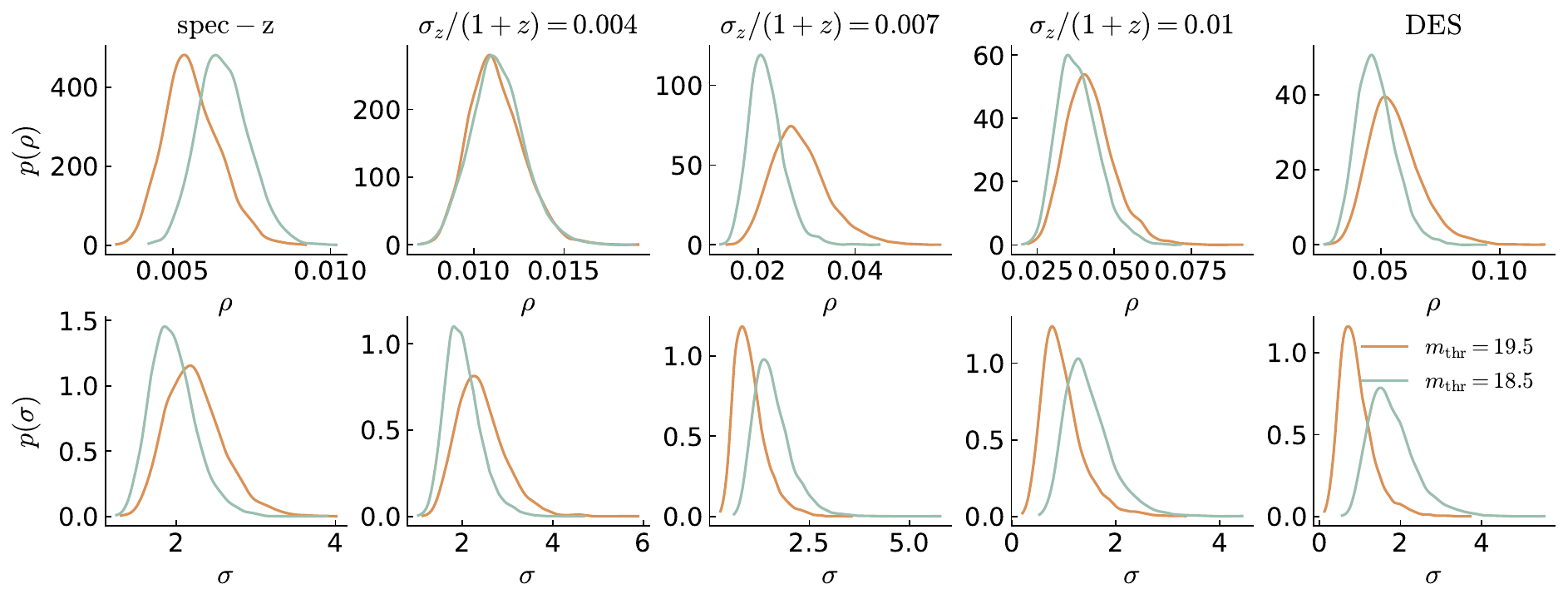}
    \caption{Posterior distribution for the GP kernel parameters under different redshift error assumptions and for the two different magnitude cuts. These results are relative to the LOS data plotted in \cref{fig:LOS-generation}.}
    \label{fig:kernel-params-constraints}
\end{figure*}

In \cref{fig:kernel-params-constraints}, we plot the posterior distribution for the GP kernel parameters obtained with the LOS shown in \cref{fig:LOS-generation}.
We note in particular that the mean value of $\rho$, defining the length scale at which the GP realizations oscillate, increases as the redshift error increases.
Physically, this means that as the redshift error increases, thereby smoothing the observed redshift distribution, the GP oscillations along redshift space are smaller and less numerous.
On the other hand, the amplitude scale parameter $\sigma$ is comparable when varying the error on galaxy redshifts.
When using the $8\%$ complete catalog ($m_{\rm thr} = 18.5$), the posteriors are usually wider than those obtained with the $24\%$ complete catalog  ($m_{\rm thr} = 19.5$).
This is expected since there is less data to constrain the GP kernel parameters.

\section{All Hubble constant constraints}\label{app:full_H0_posterior}
 In \cref{tab:H0-constraints-summary-gp} we summarize the median and the 68\% C.I. for all cases studied.
We note that all methods yield unbiased $H_0$ constraints, as the fiducial value is consistently recovered within the 68\% C.I.

\setlength\LTcapwidth{\textwidth}
\begingroup
\setlength{\tabcolsep}{4pt}\renewcommand{\arraystretch}{1.15}
\begin{longtable*}{@{}c@{\extracolsep{\fill}}ccc ccccc@{}}
\caption{Median $H_0$ values and 68\% CI (in $\si{\kilo\meter\per\second\per\mega\parsec}$) for the three cases studied: the GP-completed catalog, the standard homogeneous completion (UCV), and the complete galaxy catalog (COMP). The symmetric percentage errors are shown in parentheses.\label{tab:H0-constraints-summary-gp}}\\
\hline\hline
$m_{\rm thr}$ & $A$ & $d_L^{\rm thr}$ & Method & \texttt{spec-z} & 0.004 & 0.007 & 0.01 & \texttt{DES} \\
\hline
\endfirsthead
\multicolumn{9}{c}{TABLE~\thetable. (Continued)}\\
\hline\hline
$m_{\rm thr}$ & $A$ & $d_L^{\rm thr}$ & Method & \texttt{spec-z} & 0.004 & 0.007 & 0.01 & \texttt{DES} \\
\hline
\endhead
\hline
\multicolumn{9}{r}{\emph{(Table continued)}}\\
\endfoot
\hline\hline
\endlastfoot
18.5 & 0.1 & 1.4 & GP   & \hc{68.7}{1.4}{1.4}{2.0} & \hc{68.8}{1.4}{1.4}{2.0} & \hc{68.8}{1.4}{1.5}{2.1} & \hc{68.4}{1.6}{1.6}{2.3} & \hc{68.2}{1.9}{1.8}{2.7} \\
     &     &     & COMP & \hc{69.0}{1.5}{1.5}{2.2} & \hc{68.5}{1.7}{1.6}{2.4} & \hc{68.5}{2.0}{1.8}{2.8} & \hc{68.4}{2.4}{2.2}{3.4} & \hc{68.6}{2.8}{2.6}{3.9} \\
     &     &     & UCV  & \hc{69.1}{2.2}{2.1}{3.1} & \hc{68.4}{2.5}{2.4}{3.6} & \hc{68.1}{3.0}{2.9}{4.3} & \hc{68.1}{3.6}{3.4}{5.1} & \hc{69.0}{4.7}{4.1}{6.4} \\[0.6em]
     &     & 1.6 & GP   & \hc{69.1}{1.4}{1.4}{2.0} & \hc{68.9}{1.4}{1.3}{2.0} & \hc{68.6}{1.5}{1.4}{2.1} & \hc{69.4}{1.6}{1.5}{2.2} & \hc{70.0}{1.9}{1.8}{2.6} \\
     &     &     & COMP & \hc{69.4}{1.5}{1.4}{2.1} & \hc{69.0}{1.6}{1.6}{2.3} & \hc{68.8}{1.9}{1.7}{2.6} & \hc{68.9}{2.1}{2.1}{3.0} & \hc{68.8}{2.5}{2.3}{3.5} \\
     &     &     & UCV  & \hc{70.0}{2.6}{2.4}{3.6} & \hc{68.9}{2.8}{2.5}{3.9} & \hc{68.7}{3.4}{3.0}{4.7} & \hc{69.2}{4.1}{3.7}{5.6} & \hc{69.4}{4.9}{4.5}{6.8} \\[0.6em]
     &     & 1.8 & GP   & \hc{69.5}{1.6}{1.6}{2.3} & \hc{69.1}{1.5}{1.4}{2.1} & \hc{69.2}{1.4}{1.3}{1.9} & \hc{69.4}{1.6}{1.5}{2.2} & \hc{69.8}{1.8}{1.8}{2.6} \\
     &     &     & COMP & \hc{69.5}{1.8}{1.7}{2.5} & \hc{69.6}{1.9}{1.8}{2.7} & \hc{69.6}{2.1}{2.1}{3.0} & \hc{69.7}{2.3}{2.3}{3.3} & \hc{69.9}{2.5}{2.6}{3.6} \\
     &     &     & UCV  & \hc{68.5}{3.9}{3.8}{5.6} & \hc{68.7}{4.0}{4.1}{5.9} & \hc{69.3}{4.2}{4.4}{6.2} & \hc{70.2}{4.3}{4.8}{6.5} & \hc{71.7}{3.8}{4.9}{6.1} \\[0.6em]
     & 0.2 & 1.4 & GP   & \hc{68.6}{3.1}{3.0}{4.5} & \hc{68.2}{3.1}{2.9}{4.4} & \hc{68.3}{3.0}{3.0}{4.4} & \hc{68.4}{3.1}{3.1}{4.5} & \hc{68.2}{3.4}{3.3}{4.9} \\
     &     &     & COMP & \hc{68.1}{3.0}{2.9}{4.3} & \hc{68.0}{3.2}{3.1}{4.6} & \hc{67.8}{3.4}{3.4}{5.0} & \hc{67.2}{3.7}{3.7}{5.5} & \hc{66.2}{4.0}{4.1}{6.1} \\
     &     &     & UCV  & \hc{66.6}{3.5}{3.2}{5.0} & \hc{66.5}{3.7}{3.6}{5.5} & \hc{66.3}{4.1}{4.1}{6.2} & \hc{65.8}{4.7}{4.7}{7.1} & \hc{65.0}{5.3}{5.8}{8.5} \\[0.6em]
     &     & 1.6 & GP   & \hc{68.6}{4.0}{4.1}{5.9} & \hc{67.0}{3.6}{3.7}{5.5} & \hc{66.3}{3.5}{3.6}{5.3} & \hc{66.3}{3.5}{3.5}{5.3} & \hc{66.5}{3.5}{3.5}{5.3} \\
     &     &     & COMP & \hc{68.1}{3.9}{4.0}{5.8} & \hc{67.9}{3.9}{4.0}{5.8} & \hc{68.1}{3.7}{3.9}{5.6} & \hc{68.1}{3.7}{4.0}{5.7} & \hc{68.0}{3.7}{4.1}{5.7} \\
     &     &     & UCV  & \hc{67.8}{4.5}{4.6}{6.7} & \hc{68.0}{4.3}{4.5}{6.5} & \hc{68.5}{4.2}{4.5}{6.3} & \hc{69.1}{4.2}{4.4}{6.2} & \hc{69.9}{4.1}{4.4}{6.1} \\[0.6em]
     &     & 1.8 & GP   & \hc{70.5}{2.7}{2.9}{4.0} & \hc{69.0}{2.6}{2.8}{3.9} & \hc{67.2}{2.8}{2.9}{4.2} & \hc{67.1}{2.8}{3.0}{4.3} & \hc{67.2}{2.8}{3.2}{4.5} \\
     &     &     & COMP & \hc{68.8}{2.7}{2.9}{4.1} & \hc{68.5}{2.7}{2.9}{4.1} & \hc{68.3}{2.7}{3.0}{4.2} & \hc{67.9}{2.7}{3.1}{4.3} & \hc{67.5}{2.7}{3.1}{4.3} \\
     &     &     & UCV  & \hc{68.4}{3.0}{3.3}{4.6} & \hc{68.4}{3.0}{3.3}{4.6} & \hc{68.4}{3.0}{3.4}{4.7} & \hc{68.5}{3.0}{3.4}{4.7} & \hc{68.6}{3.1}{3.4}{4.7} \\[0.6em]
     \hline
19.5 & 0.1 & 1.4 & GP   & \hc{69.2}{1.5}{1.6}{2.2} & \hc{69.0}{1.6}{1.4}{2.2} & \hc{68.7}{1.6}{1.5}{2.3} & \hc{68.7}{1.6}{1.6}{2.3} & \hc{68.7}{1.8}{1.8}{2.6} \\
     &     &     & COMP & \hc{69.3}{1.5}{1.5}{2.2} & \hc{68.9}{1.7}{1.6}{2.4} & \hc{68.5}{2.0}{1.8}{2.8} & \hc{68.5}{2.3}{2.3}{3.4} & \hc{68.6}{2.8}{2.6}{3.9} \\
     &     &     & UCV  & \hc{69.2}{1.7}{1.7}{2.5} & \hc{68.6}{1.9}{1.8}{2.7} & \hc{68.0}{2.2}{2.1}{3.2} & \hc{67.9}{2.6}{2.5}{3.8} & \hc{68.3}{3.2}{3.0}{4.5} \\[0.6em]
     &     & 1.6 & GP   & \hc{69.3}{1.5}{1.4}{2.1} & \hc{69.2}{1.4}{1.5}{2.1} & \hc{69.0}{1.6}{1.5}{2.2} & \hc{69.3}{1.6}{1.5}{2.2} & \hc{69.7}{1.7}{1.7}{2.4} \\
     &     &     & COMP & \hc{69.5}{1.5}{1.4}{2.1} & \hc{69.1}{1.6}{1.5}{2.2} & \hc{68.9}{1.8}{1.8}{2.6} & \hc{68.9}{2.1}{2.1}{3.0} & \hc{68.8}{2.5}{2.3}{3.5} \\
     &     &     & UCV  & \hc{69.5}{1.9}{1.7}{2.6} & \hc{69.0}{2.0}{1.9}{2.8} & \hc{69.0}{2.4}{2.2}{3.3} & \hc{69.4}{2.9}{2.6}{4.0} & \hc{69.9}{3.4}{3.1}{4.7} \\[0.6em]
     &     & 1.8 & GP   & \hc{69.2}{1.7}{1.7}{2.5} & \hc{69.3}{1.7}{1.7}{2.5} & \hc{69.3}{1.7}{1.7}{2.5} & \hc{69.2}{1.8}{1.8}{2.6} & \hc{69.1}{1.8}{1.8}{2.6} \\
     &     &     & COMP & \hc{69.2}{1.8}{1.8}{2.6} & \hc{69.4}{1.9}{1.9}{2.7} & \hc{69.6}{2.1}{2.1}{3.0} & \hc{69.7}{2.3}{2.3}{3.3} & \hc{69.9}{2.5}{2.6}{3.6} \\
     &     &     & UCV  & \hc{67.9}{2.5}{2.4}{3.6} & \hc{67.9}{2.8}{2.7}{4.0} & \hc{68.2}{3.1}{3.0}{4.5} & \hc{68.0}{3.7}{3.4}{5.2} & \hc{68.4}{4.1}{3.9}{5.8} \\[0.6em]
     & 0.2 & 1.4 & GP   & \hc{68.5}{3.1}{3.0}{4.5} & \hc{68.5}{3.1}{3.0}{4.5} & \hc{68.4}{3.2}{3.0}{4.5} & \hc{68.6}{3.2}{3.2}{4.7} & \hc{68.8}{3.3}{3.4}{4.9} \\
     &     &     & COMP & \hc{68.3}{3.0}{2.9}{4.3} & \hc{68.1}{3.2}{3.1}{4.6} & \hc{67.8}{3.4}{3.4}{5.0} & \hc{67.2}{3.6}{3.7}{5.4} & \hc{66.2}{4.0}{4.1}{6.1} \\
     &     &     & UCV  & \hc{67.9}{3.2}{3.0}{4.6} & \hc{68.0}{3.4}{3.2}{4.8} & \hc{67.7}{3.7}{3.6}{5.4} & \hc{67.4}{4.0}{4.0}{5.9} & \hc{66.7}{4.4}{4.4}{6.6} \\[0.6em]
     &     & 1.6 & GP   & \hc{68.2}{3.7}{3.8}{5.5} & \hc{67.9}{3.7}{3.9}{5.6} & \hc{68.0}{3.7}{3.7}{5.4} & \hc{68.5}{3.5}{3.7}{5.3} & \hc{69.1}{3.4}{3.6}{5.1} \\
     &     &     & COMP & \hc{68.1}{3.9}{3.9}{5.7} & \hc{68.0}{3.7}{4.0}{5.7} & \hc{68.1}{3.7}{3.9}{5.6} & \hc{68.1}{3.7}{4.0}{5.7} & \hc{68.0}{3.7}{4.1}{5.7} \\
     &     &     & UCV  & \hc{68.2}{4.1}{4.1}{6.0} & \hc{68.0}{4.0}{4.1}{6.0} & \hc{68.6}{4.0}{4.0}{5.8} & \hc{69.2}{4.0}{4.2}{5.9} & \hc{69.9}{3.9}{4.3}{5.9} \\[0.6em]
     &     & 1.8 & GP   & \hc{68.6}{2.6}{2.9}{4.0} & \hc{68.8}{2.7}{2.8}{4.0} & \hc{68.7}{2.6}{2.9}{4.0} & \hc{68.7}{2.6}{2.9}{4.0} & \hc{68.6}{2.7}{2.9}{4.1} \\
     &     &     & COMP & \hc{68.6}{2.7}{3.0}{4.2} & \hc{68.4}{2.7}{2.9}{4.1} & \hc{68.3}{2.7}{3.0}{4.2} & \hc{67.9}{2.7}{3.1}{4.3} & \hc{67.5}{2.7}{3.1}{4.3} \\
     &     &     & UCV  & \hc{68.5}{2.8}{3.1}{4.3} & \hc{68.5}{2.9}{3.1}{4.4} & \hc{68.6}{2.9}{3.2}{4.5} & \hc{68.6}{2.9}{3.3}{4.5} & \hc{68.5}{3.0}{3.3}{4.6} \\
\end{longtable*}
\endgroup

\bibliographystyle{apsrev4-2}
\bibliography{bibliography}

\end{document}